\documentclass[aps,prb,twocolumn,showpacs,floatfix,superscriptaddress]{revtex4-1}
\usepackage{graphicx}
\usepackage{url}
\usepackage{hyperref}
\usepackage{inconsolata}
\usepackage{amsmath}

\usepackage{braket}
\usepackage{xcolor} %

\begin{document}
\title{Field-driven quantum phase transitions in $S=\frac{1}{2}$ spin chains}

\author{Adam Iaizzi}
\email{iaizzi@bu.edu}
\affiliation{Department of Physics, Boston University, 590 Commonwealth Avenue, Boston, Massachusetts 02215, USA}

\author{Kedar Damle}
\affiliation{Department of Theoretical Physics, Tata Institute of Fundamental Research, Mumbai 400 005, India}

\author{Anders W. Sandvik}
\affiliation{Department of Physics, Boston University, 590 Commonwealth Avenue, Boston, Massachusetts 02215, USA}

\date{\today}

\begin{abstract}
We study the magnetization process of a one-dimensional extended Heisenberg model, the \mbox{$J$-$Q$} model, as a function of an external magnetic field, $h$. 
In this model, $J$ represents the traditional antiferromagnetic Heisenberg exchange and $Q$ is the strength of a competing four-spin interaction. 
Without external field, this system hosts a twofold-degenerate dimerized (valence-bond solid) state above a critical value $q_c\approx 0.85$ where $q\equiv Q/J$. 
The dimer order is destroyed and replaced by a partially polarized translationally invariant state at a critical field value. 
We find magnetization jumps (metamagnetism) between the partially polarized and fully polarized state for $q >q_{\rm min}$, where we have calculated $q_{\rm min}=\frac{2}{9}$ exactly. 
For $q>q_{\rm min}$ two magnons (flipped spins on a fully polarized background) attract and form a bound state. 
Quantum Monte Carlo studies confirm that the bound state corresponds to the first step of an instability leading to a finite magnetization jump for $q > q_{\rm min}$. 
Our results show that neither geometric frustration nor spin-anisotropy are necessary conditions for metamagnetism. 
Working in the two-magnon subspace, we also find evidence pointing to the existence of metamagnetism in the unfrustrated \mbox{$J_1$-$J_2$} chain ($J_1>0$, $J_2<0$), but only if $J_2$ is spin-anisotropic. 
In addition to the studies at zero temperature, we also investigate quantum-critical scaling near the transition into the fully polarized state for $q \le q_{\rm min}$ at $T>0$. 
While the expected ``zero-scale-factor'' universality is clearly seen for $q=0$ and $q \ll q_{\rm min}$, for $q$ closer to $q_{\rm min}$ we find that extremely low temperatures are required to observe the asymptotic behavior, due to the influence of the tricritical point at $q_{\rm min}$. 
In the low-energy theory, one can expect the quartic nonlinearity to vanish at $q_{\rm min}$ and a marginal sixth-order term should govern the scaling, which leads to a cross-over at a temperature $T^*(q)$ between logarithmic tricritical scaling and zero-scale-factor universality, with $T^*(q) \to 0$ when $q \to q_{\rm min}$.
\end{abstract}

\maketitle

\section{Introduction}

In this paper we characterize the magnetization process of a one-dimensional Heisenberg antiferromagnet with four-spin interactions of strength $Q$ in addition to the standard antiferromagnetic exchange term of strength $J$ (the \mbox{$J$-$Q$} model \cite{sandvik2007,sandvik2011computational}) as it is subjected to an external magnetic (Zeeman) field. 
The model is defined in terms of singlet projectors acting on a lattice of $S=1/2$ sites: 
\begin{equation}
P_{i,j} \equiv \frac{1}{4} - {\bf S}_i \cdot {\bf S}_j.
\end{equation}
The standard antiferromagnetic Heisenberg exchange is equivalent to $-JP_{ij}$ with $J>0$. 
In the \mbox{$J$-$Q$} model this interaction is supplemented by the product $-QP_{i,j}P_{k,l}$ (or products of more than two projectors \cite{kaul2013}) with the site pairs $i,j$ and $k,l$ suitably arranged and summed over the lattice sites with all lattice symmetries respected. 
The long-range ordered (in two or three dimensions) or critical (in one dimension) antiferromagnetic (AFM) state of the pure Heisenberg model can be destroyed for sufficiently large $Q/J$. 
A non-magnetic ground state with broken lattice symmetries due to dimerization (a valence-bond solid, VBS) then appears. 
The VBS state and the quantum phase transition between the AFM and VBS states have been studied extensively in both one \cite{sanyal2011,tang2011a,tang2014} and two \cite{sandvik2007,lou2009,sandvik2010,jin2013,tang2013} dimensions. 
The \mbox{$J$-$Q$} model is a member of a broad family\cite{kaul2013} of Marshall-positive spin Hamiltonians constructed from products of any number of singlet projection and permutation operators. 

Here we consider the simplest one-dimensional (1D) \mbox{$J$-$Q$} model, where the $Q$ term is composed of a product of just two singlet projection operators:
\begin{equation}
H_{JQ} = -J \sum \limits_i P_{i,i+1} - Q \sum \limits_{i} P_{i,i+1} P_{i+2,i+3}, \label{eq:jq2}
\end{equation}
and add an external magnetic field of strength $h_z$ to define the \mbox{$J$-$Q$-$h$} model:
\begin{equation}
H_{JQh} = H_{JQ} -h_z \sum \limits_i S^z_i. \label{eq:jq2h}
\end{equation}
We set the energy scale by fixing $J=1$ and refer to the dimensionless parameters $q\equiv Q/J$ and $h \equiv h_z/J$. 

Our focus will be on the magnetization curve as a function of the field, which we study both at $T=0$ and $T>0$. 
We use the stochastic series expansion (SSE) \cite{sandvik1991,sandvik2011computational} quantum Monte Carlo (QMC) method with directed loop updates, \cite{sandvik_dl} supplemented by quantum replica exchange \cite{hukushima1996,sengupta2002} to alleviate metastability problems in the simulations.
We show that the $Q$ term has dramatic consequences for the magnetization process. 
In the pure Heisenberg chain ($q=0$), and for small $q$, the magnetization curve at temperature $T=0$ is continuous.
When $q$ exceeds a critical value, a magnetization jump (metamagnetic transition) \cite{jacobs1967,stryjewski1977} appears between a partially magnetized and the fully polarized state. 
Using an ansatz motivated by numerical results for two magnons in a saturated background, we obtain an exact analytical result for the minimum coupling ratio, $q_{\rm min}$, at which such a magnetization jump can occur; $q_{\rm min}=\frac{2}{9}$. 
This calculation also reveals the mechanism of the magnetization jump: the onset of attractive magnon interactions when $q>q_{\rm min}$. 
At exactly $q_{\rm min}$, the magnons behave as effectively non-interacting particles. 
The onset of a bound state of magnons is a general mechanism for metamagnetism,\cite{aligia2000,arlego2011} but normally this phenomenon has been associated with frustration due to competing exchange couplings \cite{gerhardt1998,hirata1999,aligia2000,dmitriev2006,kecke2007,sudan2009,arlego2011,kolezhuk2012} or strong spin anisotropy \cite{gerhardt1998,hirata1999,aligia2000} [including the classical two-dimensional (2D) Ising model with second-neighbor interactions\cite{landau1972,rikvold1983}].
We believe this effect could also explain the metamagnetic transition reported in a ring exchange model, \cite{huerga2014} (a close relative of the \mbox{$J$-$Q$} model), where the metamagnetic transition corresponds to a first-order transition from a partially occupied to a fully occupied state. 
Our study provides an example of metamagnetism in a spin-isotropic system without traditional frustration. Note that the onset value $q_{\rm min}=\frac{2}{9}$ of metamagnetism is much smaller than the critical value $q_c \approx 0.85$ at which the chain dimerizes in the absence of a field.
Thus, the metamagnetism here is not directly related to the VBS state of the \mbox{$J$-$Q$} model.

A bound state of magnons \textit{does not} occur in the standard \mbox{$J_1$-$J_2$} Heisenberg chain \cite{majumdar1,majumdar2,soos2016} with frustrated antiferromagnetic couplings $J_1>0$, $J_2>0$, but it \textit{does} occur \cite{dmitriev2006,sudan2009,arlego2011} for the also-frustrated FM-AFM regime $J_1<0$, $J_2>0$. 
In our study of the unfrustrated regime, we find bound magnon states in the \mbox{$J_1$-$J_2$} chain with a ferromagnetic (FM) second-neighbor coupling (AFM $J_1>0$, FM $J_2<0$), but only if this second-neighbor coupling is also spin anisotropic, of the form $J_2[S^z_iS^z_j + \Delta (S^x_iS^x_j+S^y_iS^y_j)]$. 
The existence of a bound state for some values of the parameters $\Delta \neq 0$ and $|J_2/J_1|$ is likely a precursor to a metamagnetic transition as in the \mbox{$J$-$Q$-$h$} chain, but we do not study it further with QMC here.

We also study the \mbox{$J$-$Q$-$h$} chain at $T>0$ in the region close to magnetic saturation when $q \le q_{\rm min}$. 
Here one would expect the dependence of the magnetization on the field and the temperature to be governed by a remarkably simple ``zero-scale-factor''
universal critical scaling form. \cite{sachdev1994} 
We observe this behavior clearly for $q=0$ and $q \ll q_{\rm min}$.
For $q$ closer to $q_{\rm min}$ we find that the scaling form is only obeyed at extremely low temperatures, due to onset of metamagnetism at $q=q_{\rm min}$. 
We expect $q_{\rm min}$ to be a tricritical point at which the sign of the quartic coupling ($|\psi|^4$) of the boson field changes in the low-energy effective field theory of the system. 
This corresponds to the two-magnon interaction switching from repulsive to attractive at this point. 
Precisely at $q=q_{\rm min}$, the two-magnon interaction vanishes and the system is dominated by three-body interactions, represented in the effective field theory by a $|\psi|^6$ term which is marginal in $d=1$. 
The smallness of the quartic term close to $q_{\rm min}$ leads to a cross-over, which we observe, between tricritical and zero-scale-factor behavior, with the cross-over temperature approaching zero as $q \to q_{\rm min}$.

The outline of the rest of the paper is as follows: In Sec.~\ref{s:methods} we briefly summarize the numerical methods we have used. 
We then discuss the phase diagram of the \mbox{$J$-$Q$-$h$} model in Sec. \ref{s:pd}. 
In Secs. \ref{s:jump} and \ref{s:j1j2} we discuss metamagnetism in the \mbox{$J$-$Q$-$h$} and \mbox{$J_1$-$J_2$} chains, respectively. Section \ref{s:zsf} contains our results for zero-factor scaling of the saturation transition in the \mbox{$J$-$Q$-$h$} chain. 
In Sec.~\ref{s:discussion} we summarize and discuss our main results.

\section{Methods \label{s:methods}}

The primary numerical tools employed in this work are Lanczos exact diagonalization and the SSE QMC method \cite{sandvik1991} with directed loop updates. \cite{sandvik_dl} 
Symmetries are implemented in the Lanczos calculations as described in Ref.~\onlinecite{sandvik2011computational}.
SSE works by exactly mapping a \mbox{$d$-dimensional} quantum problem onto a \mbox{$(d+1)$-dimensional} classical problem through Taylor expansion of ${\rm e}^{-\beta H}$. 
This extra dimension is related to imaginary time in a manner similar to the path integrals in world-line QMC, but in the Monte Carlo sampling the operational emphasis is not on the paths but on the operators determining the fluctuations of the paths. 
We incorporate the magnetic field in the diagonal part of the two-spin ($J$) operators. 
Diagonal updates insert and remove two- and four-spin diagonal operators, while the directed loop updates change the operators from diagonal to off-diagonal and vice-versa. \cite{sandvik2011computational} 
When a two-spin operator is encountered in the loop-building process, we choose the exit leg using the ``no-bounce'' solution of the directed loop equations for the Heisenberg model in an external field found in Ref.~\onlinecite{sandvik_dl}. 
When encountering a four-spin \mbox{$Q$-type} operator, where the field contribution is not present, the exit leg is chosen using a deterministic ``switch and reverse" strategy, essentially identical to the SSE scheme for the standard isotropic Heisenberg model. \cite{sandvik2011computational}

When using SSE alone, we found that simulations sometimes became stuck at metastable magnetization values for long periods of time. 
This made it hard for simulations to reach equilibrium and difficult to compute accurate estimates of statistical errors. 
This problem can be easily seen in our preliminary results presented in Figs. 2 and 3 of Ref.~\onlinecite{iaizzi2015}, where the large fluctuations in the magnetization are due to this `sticking' problem. 
To remedy this, in the present work we implemented a variation of the replica exchange method \cite{hukushima1996} for QMC known as quantum replica exchange, \cite{sengupta2002} implemented using the MPI (Message Passing Interface) parallel computing library.

In the traditional replica exchange method \cite{hukushima1996} (also known as parallel tempering), many simulations are run in parallel on a mesh of temperatures. 
In addition to standard Monte Carlo updates, replicas are allowed to swap temperatures with each other with some probability that preserves detailed balance in the extended multi-canonical ensemble. 
This allows a replica in a metastable state to escape by wandering to a higher temperature. 
In the SSE simulations with replica exchange, \cite{sengupta2002} we run many ($10\sim100$) simulations in parallel. 
Instead of using different temperatures as in standard parallel tempering, we use a mesh of magnetic fields.  
After each Monte Carlo sweep, we allow replicas to exchange magnetic fields with one another in a manner that preserves detailed balance within the ensemble of SSE configurations. 

For relatively little communications overhead, we find that replica exchange can dramatically reduce equilibration and autocorrelation times, thus allowing simulations of much larger systems at much lower temperatures. 
In practice, adding additional replicas slows down the simulation because the time required to complete a Monte Carlo sweep varies and all the replicas have to wait for the slowest replica to finish before continuing. 
This slowdown can be somewhat alleviated by running more than one replica on each core. 

\section{Phase Diagram \label{s:pd}}

\begin{figure}
\includegraphics[width=62mm]{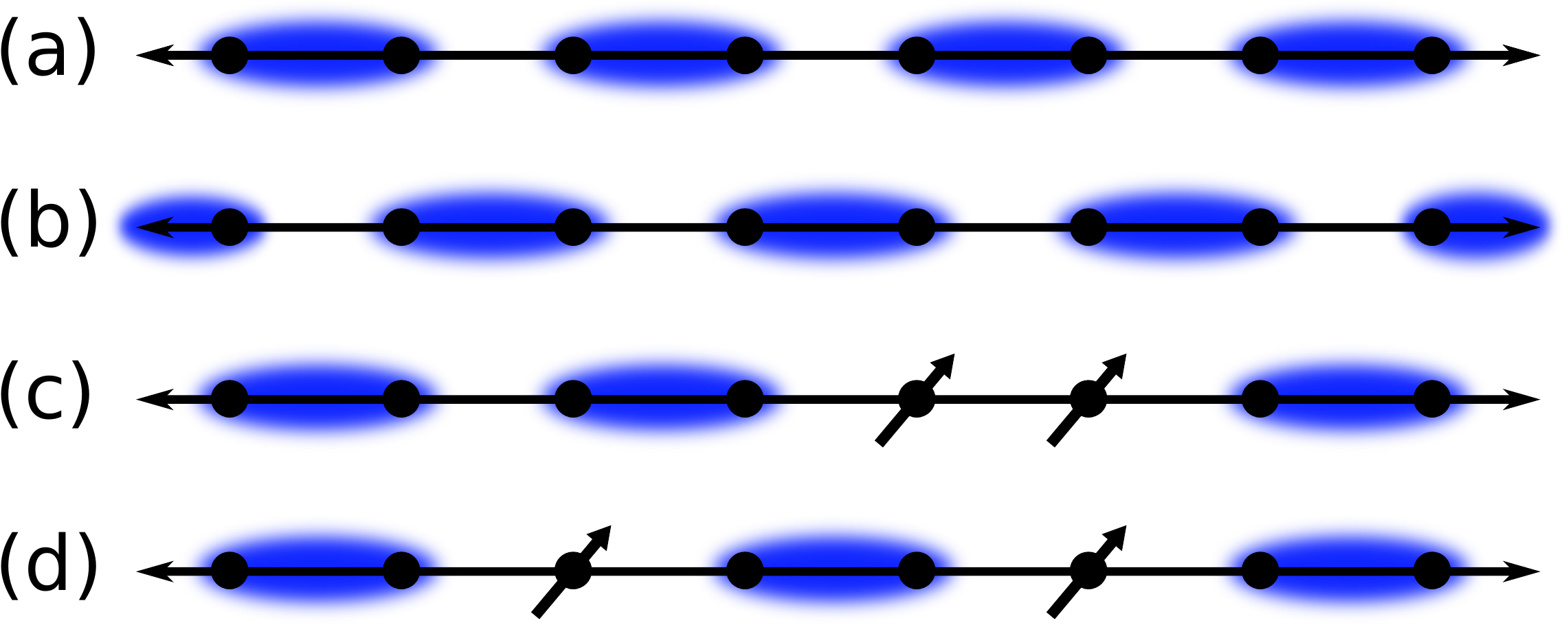}
\caption{Examples of VBS configurations of $S={1}/{2}$ spins in one dimension. Each blue ellipse represents a singlet pair: $\left(\ket{\uparrow \downarrow}-\ket{\downarrow \uparrow }\right)/\sqrt 2$ . In a real VBS there are also fluctuations in the singlet patterns (except in special cases) but the density of singlets on the bonds is still modulated with periodicity two lattice spacings. (a), (b) Show the two degenerate VBS ground states, (c) illustrates a triplet excitation in which a singlet bond is broken, and (d) illustrates a triplet excitation deconfined into two independently propagating spinons. \label{vbs}}
\end{figure}

The \mbox{$J$-$Q$} model has so far been of theoretical interest mainly as a tool for large-scale studies of VBS phases and AFM--VBS transitions. 
In a VBS (dimerized state), spins pair up to form a crystal of localized singlets, thus breaking translational symmetry but preserving spin-rotation symmetry as illustrated in Figs.~\ref{vbs}(a) and \ref{vbs}(b). 
The elementary quasiparticle excitations of a VBS are gapped triplet waves (triplons) formed by exciting a singlet pair to a triplet, as seen in Fig.~\ref{vbs}(c). 
Triplons sometimes deconfine into pairs of spinons: fractionalized spin-${1}/{2}$ excitations that correspond to VBS domain walls as shown in Fig.~\ref{vbs}(d). 
For dimensionality $d>1$, the spinons are confined by a string in a manner similar to quarks, the energy associated with the shifted VBS arrangement resulting from separating two spinons is directly proportional to the distance between the spinons (see Ref.~\onlinecite{deconfwall} for a recent discussion of this analogy). 
In a one-dimensional VBS, the spinons are always deconfined, unless the Hamiltonian breaks translational symmetry. \cite{haldane1982,tang2011a} 
The frustrated Hamiltonians that were traditionally used to study VBS physics, e.g., the \mbox{$J_1$-$J_2$} chain, \cite{majumdar1,majumdar2,majumdar1970,haldane1982} suffer from the sign problem, which prevents large-scale numerical simulations using QMC methods; the \mbox{$J$-$Q$} model is sign-problem free. 

Our main aim in this paper is to study the magnetization process of the \mbox{$J$-$Q$-$h$} chain from $h=0$ all the way to the fully polarized state where the concept of spinons in a dimer background breaks down.
To understand the basic physics in this regime, it is more appropriate to consider flipped spins (``magnons'') relative to the vacuum of a fully magnetized state. 
For completeness, in this section we also comment on the $T=0$ phases of the system in the full $q$-$h$ plane.

\begin{figure}
\includegraphics[width=68mm]{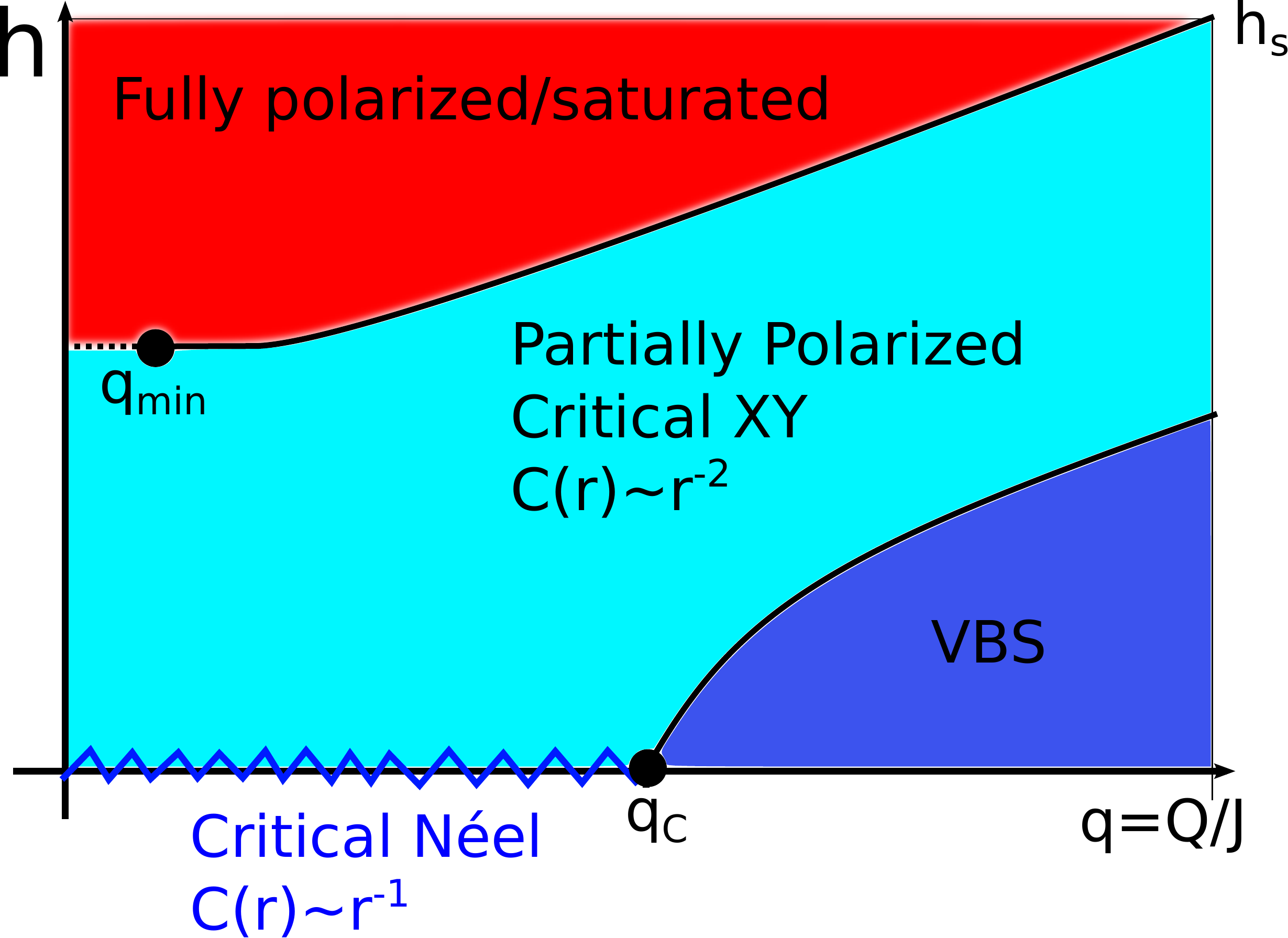}
\caption{Schematic phase diagram of the \mbox{$J$-$Q$-$h$} chain defined in Eqs.~(\ref{eq:jq2}) and (\ref{eq:jq2h}). The different phases and special points indicated are described in the text.
\label{pd}}
\end{figure}

Figure \ref{pd} shows a schematic phase diagram assembled from the literature and  our own calculations. 
The parameter regions corresponding to the horizontal and vertical axes are well understood from past studies; the off-axes area has not been previously studied and is therefore the primary focus of this paper. 
The $h$ axis is the standard Heisenberg chain in a magnetic field, where the transition into the fully polarized state is continuous. 
The $q$ axis corresponds to the previously-studied zero-field \mbox{$J$-$Q$} model, \cite{tang2011a} where for $q<q_c$ there is a Heisenberg-type critical AFM state with spin-spin correlations decaying with distance $r$ as $1/r$ (up to a multiplicative logarithm). \cite{singh1989} 
At \mbox{$q=q_c \approx 0.8483$} the chain undergoes a dimerization transition into a VBS ground state. \cite{tang2011a} 
This transition is similar to the Kosterlitz-Thouless transition and identical to the quasi-AFM to VBS transition in the \mbox{$J_1$-$J_2$} chain. \cite{haldane1982,tang2011a,sanyal2011}

In the full phase diagram for $q>0$ (which we focus on here because $q<0$ leads to QMC sign problems), there are three phases: a fully polarized phase, a VBS, and a partially polarized critical XY phase. 
If we start from a VBS state ($h=0,q>q_c$) and add a magnetic field, the field will `pull down' the triplet excitations with magnetization $m_z>0$ and at some $h_c(q)$ a magnetized state becomes the ground state.
These triplets originating from ``broken singlets'' will deconfine into spinons,\cite{tang2011a,mourigal2013} as illustrated in Figs.~\ref{vbs}(c) and \ref{vbs}(d).
Each spinon constitutes a domain wall between VBS-ordered domains (as discussed in detail in Ref.~\onlinecite{tang2011a}), and we therefore expect any finite density of spinons to destroy the VBS order. 
The phase boundary extending from $q_c$ should therefore follow the gap to excite a single triplet out of the VBS. 
We expect the destruction of the VBS to yield a partially polarized state with critical XY correlations, as in the standard AFM Heisenberg chain in an external field. 
We do not focus on this part of the phase diagram here, and will not discuss the nature of the VBS--XY transition or the exact form of this phase boundary. 

We focus mainly on the line $h_s(q)$ separating the XY and saturated phases in Fig.~\ref{pd}, and will provide quantitative results in the following sections. 
The magnetization curve is continuous along the dotted portion of $h_s$; here, the saturation transition is governed by a remarkably simple zero-scale-factor universality. \cite{sachdev1994} 
The solid portion denotes the presence of a magnetization jump: a first-order quantum phase transition known as the metamagnetic transition. 
The point $q_{\rm min}$ marks the lower metamagnetic bound, a tricritical point where the magnetization jump is infinitesimal.

\section{Metamagnetism in the \mbox{$J$-$Q$} chain\label{s:jump}}

\begin{figure}
\includegraphics[width=75mm]{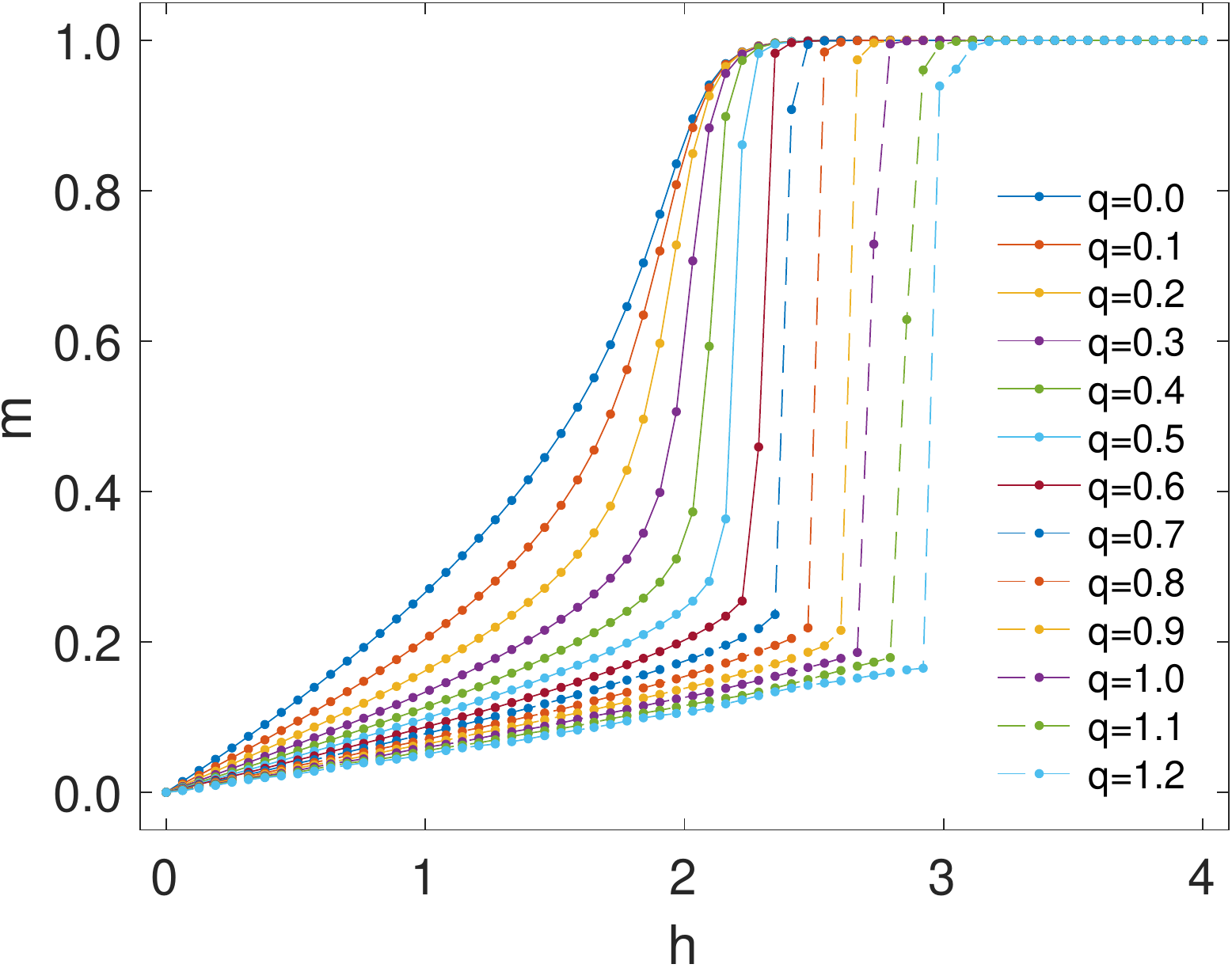}
\caption{Magnetization density of the \mbox{$J$-$Q$-$h$} chain as a function of the external field for a set of coupling ratios $0 \leq q \leq 1.2$ (from Heisenberg limit to beyond the VBS transition). The system size is $L=96$ and the inverse temperature is $\beta = 12$ in all cases. Error bars are smaller than the markers. \label{varq}}
\end{figure}

The introduction of the four-spin $Q$ term has a dramatic effect on the magnetization process. 
In Fig. \ref{varq}, we plot the magnetization density, $m(h)$, normalized to be unity in the fully polarized state, 
\begin{align}
m \equiv \frac{2}{L} \sum \limits_{i=1}^L \Braket{S^z_i}, \label{eq:magden}
\end{align}
for periodic \mbox{$J$-$Q$-$h$} chains with $0 \leq q \leq 1.2$, $L=96$, and inverse temperature $\beta=12$ (where the finite-temperature effects are already
small on the scale used in the figure). 
We begin in the Heisenberg limit ($q=0$) and increase $q$. For small $q$, the saturation field is unchanged, but the shape of the magnetization curve changes significantly, becoming steeper near saturation. 
As $q$ increases, the magnetization seems to develop a jump to saturation and the size of this jump grows with increasing $q$. 
It is especially interesting that this jump appears for $q<q_c$, a regime where the $h=0$ chain is in the critical AFM state and not yet in the VBS state. 
This magnetization jump is an example of a metamagnetic transition \cite{jacobs1967,stryjewski1977} and shows many hallmarks of a first-order phase transition, including hysteresis in the QMC simulations (as documented in our earlier, preliminary paper \cite{iaizzi2015}). 

\begin{figure}
\includegraphics[width=75mm]{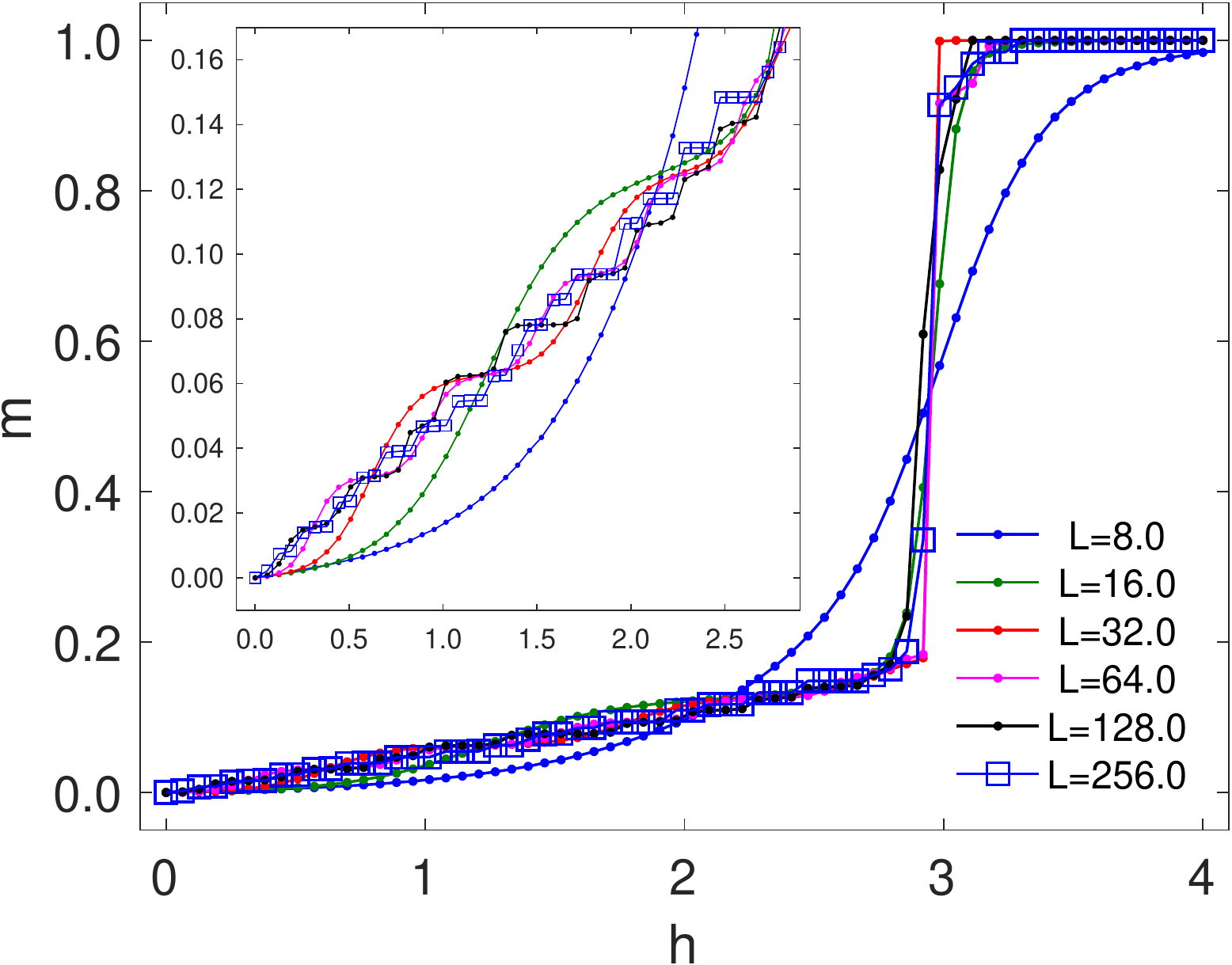}
\caption{Magnetization density of the \mbox{$J$-$Q$-$h$} chain at $q=1.2$ as a function of the external field $h$, with the inverse temperature scaled with size as $\beta=L/4$. The system sizes are between $L=8$ and $256$ as indicated. The inset shows a zoomed-in view of the paramagnetic regime. The error bars are smaller than the markers in main figure and have been omitted for clarity also in the inset (where they are some times slightly larger than the markers).  \label{magvh}}
\end{figure}

In Fig. \ref{magvh} we plot the magnetization density at \mbox{$q=1.2$} for chains of sizes ranging from $L=8$ to $256$ and inverse temperature $\beta=L/4$. 
In this regime, we observe two distinct phases: a paramagnetic regime and a fully polarized state separated by a sharp jump. 
The magnetization curves exhibit near perfect agreement for all sizes studied, limited only by the discretized values of $m$ for each size (visible in greater detail in the inset). Because of the way in which the temperature is scaled, for the smallest sizes the steps are thermally smeared out but become visible for the longer chains. 
Figure~\ref{magvh}, as in Fig.~\ref{varq}, shows no signs of any magnetization plateaus apart from the fully polarized one. 
There is also no sign of the VBS gap (to the first triplet excitation), which should manifest itself as a $m=0$ plateau for $q>q_c$, reflecting the finite field needed to close the gap. While there \textit{is} a gap in the VBS, at these sizes and temperatures the VBS gap is too small to produce a noticeable effect. 
We have computed finite-size gaps using Lanczos calculations but they are difficult to extrapolate to infinite size, and we can only extract an upper bound; the
triplet gap at $q=1.2$ should be less than $0.02$. \cite{sylvain2017}

\begin{figure}
\includegraphics[width=75mm]{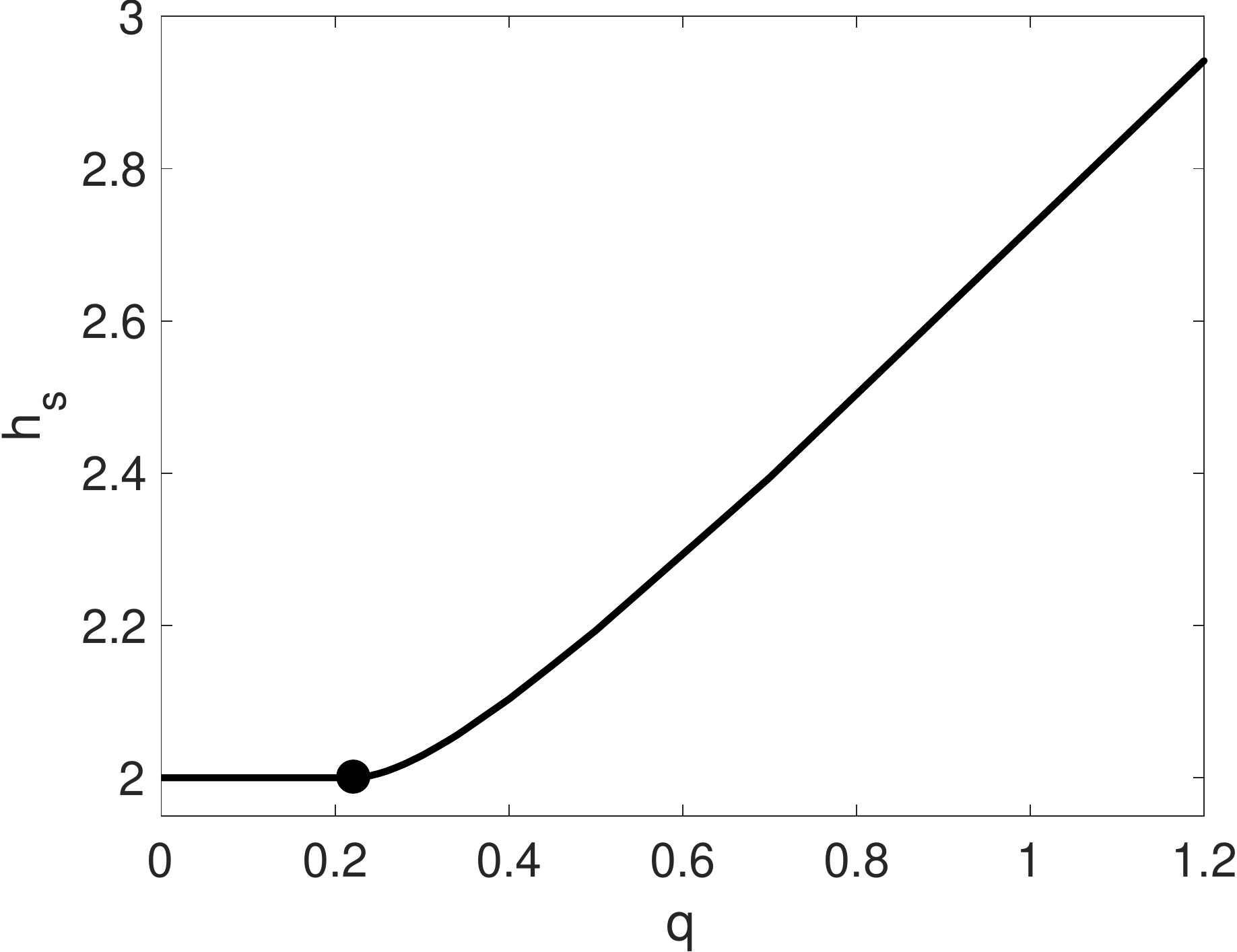}
\caption{Saturation field versus the coupling ratio for the $L=30$ periodic \mbox{$J$-$Q$-$h$} chain calculated using the Lanczos method. The dot indicates $q_{\rm min}$. \label{hcrit}}
\end{figure}

It was difficult to extract precise results for the saturation field $h_s$ or $m_c$ (the magnetization at which the jump occurs) due to the tendency of simulations to get stuck in metastable states near the transition \cite{iaizzi2015} (itself a characteristic of a first-order transition). 
Although the use of replica exchange has dramatically reduced this problem, it is still apparent for large chains and at lower temperatures. 
To extract $h_s$ precisely, we therefore used Lanczos exact diagonalization. 
The external field commutes with the Hamiltonian, so we can diagonalize the zero-field \mbox{$J$-$Q$} model and add the contribution from the field in afterwards. 
Figure~\ref{hcrit} shows the critical magnetic field for $L=30$ (we have also studied smaller systems in this way). 
For $q \leq q_{\rm min}$, the saturation field is exactly $h_s=2J$. 
In this regime, $h_s$ is determined by a level crossing between the $m=S$ and $S-1$ states which is independent of both $q$ and $L$; see also Eqs.~(\ref{eq:e1}) and (\ref{eq:energies}a) and (\ref{eq:energies}b). 
For $q > q_{\rm min}$, we find a positive relationship between $h_s$ and $q$, consistent with our QMC results in Fig.~\ref{varq}; here we should expect some finite-size effects, but they do not alter the qualitative character of the line $h_s(q)$. 

\subsection{Origin of the magnetization jump\label{s:qmin}}

Although the excitations of the zero-field \mbox{$J$-$Q$} chain are classified as spinons, near the saturation transition the density of domain walls is too high for this picture to be relevant, and the excitations are better characterized as magnons: bosonic spin-1 excitations corresponding to spin flips on a background of uniformly polarized spins. 
We will now show that the magnetization jump in the \mbox{$J$-$Q$-$h$} chain (and later, the \mbox{$J_1$-$J_2$} chain with anisotropy in Sec.~\ref{s:j1j2}) is caused by the onset of an effective attractive interaction between these magnons.

Using an analytical approach and diagonalization of short chains, we will now derive $q_{\rm min}$, the minimum value of $q$ required to produce a jump (see Fig.~\ref{pd}). 
This argument is described in more detail in Appendix~\ref{s:ajq}. 
We begin with the fact that the jump is always to the saturated state and assume that the size of the jump \mbox{$\Delta m_z/L \rightarrow 0$} at $q_{\rm min}$ as \mbox{$L \rightarrow \infty$}. 
In an infinite system, the smallest possible jump is infinitesimal; in this case the ``jump'' corresponds only to a higher-order singularity (a divergence of the magnetic susceptibility). 
In a finite-size system, the magnetization advances by steps of \mbox{$\Delta m_z \geq 1$}. 
In a trivial paramagnet, the magnetization advances by the smallest possible increment: \mbox{$\Delta m_z = 1$}; this effect can be seen for $L=256$ in the inset of Fig.~\ref{magvh}. 
Larger magnetization steps indicate the presence of some nontrivial effect; the smallest \textit{nontrivial} jump is $\Delta m_z = 2$,  i.e., a direct level crossing between $m_z = S-2$ and $S$. 
In Appendix \ref{s:ajq}, we discuss the details of a two-magnon approach to solving this problem using the condition for the level crossing: 
\begin{align}
\bar E_2 \leq 2 \bar E_1 \label{eq:jumpCondition1},
\end{align}
where $\bar E_n$ is the zero-field $n$-magnon ground-state energy as defined in Eq.~(\ref{eq:energies}). 

Equation~(\ref{eq:jumpCondition1}) essentially requires that the interaction between the magnons be attractive, since the energy of two interacting magnons is \textit{lower} than twice the energy of a single magnon. 
Metamagnetism can be brought on by the appearance of bound states of magnons if there is an instability toward bound states of ever more magnons. 
Thus, the existence of such a bound state is suggestive of, but does not guarantee, the existence of a macroscopic magnetization jump. 
If the bound pairs of magnons are not attracted to other bound pairs of magnons, then the magnetization merely advances by steps of $\Delta m_z=2$ without any macroscopic jump. 
This effect has been documented previously:\cite{honecker2006,kecke2007} in a liquid of bound states of two or more magnons, the magnetization undergoes microscopic jumps  where $\Delta m_z$ is an integer equal to the number of bound magnons, with in principle, an infinite number of such phases existing, but never a macroscopic jump. 

\begin{figure}
\includegraphics[width=75mm]{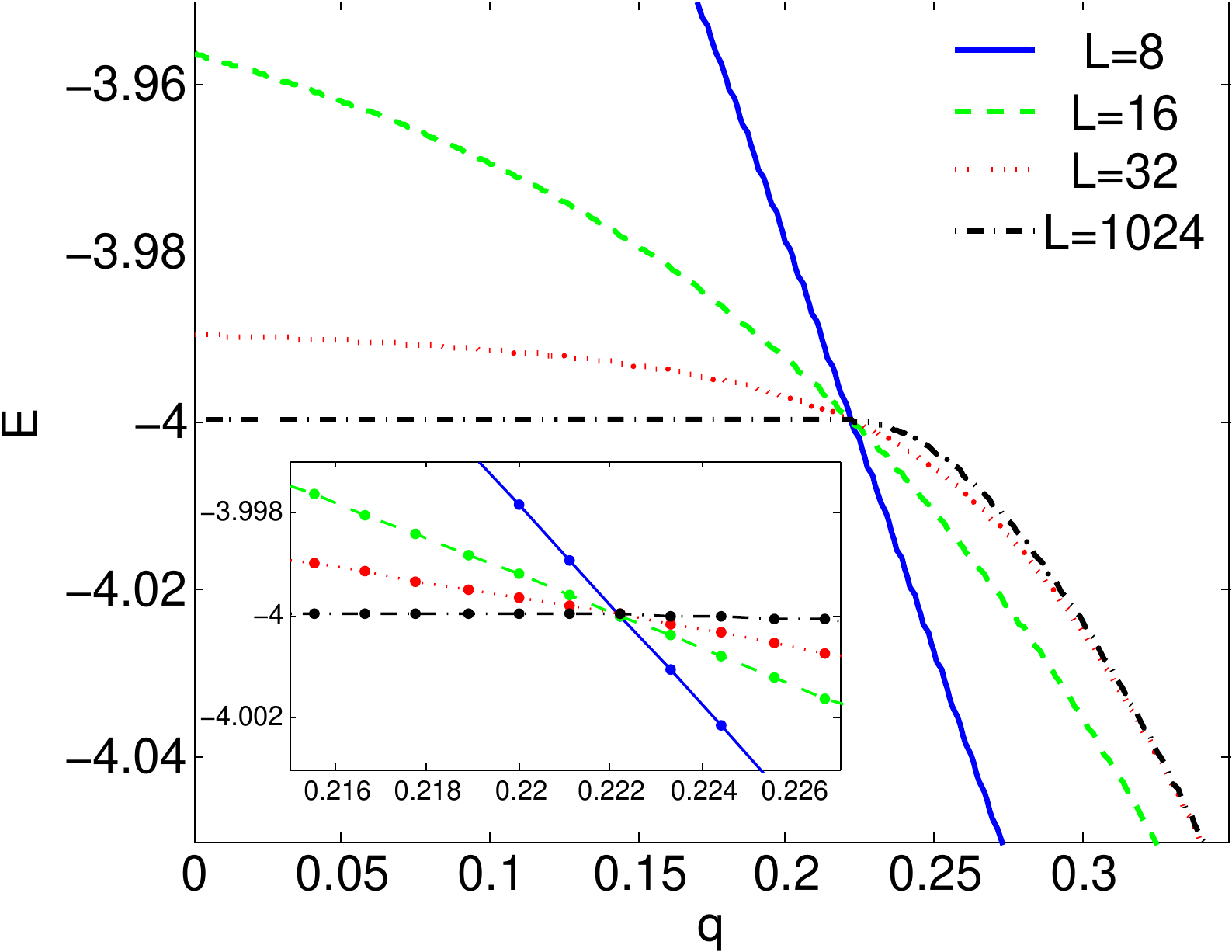}
\caption{The lowest-energy eigenvalue \mbox{$\bar E_2 (J=1,Q=q,L)$} in the two-magnon sector \mbox{($m_z = S-2$)} in the \mbox{$J$-$Q$-$h$} chain for system sizes \mbox{$L=8,16,32,1024$}. \label{e2}}
\end{figure}

Thanks to the QMC data, there can be no doubt of the existence of a macroscopic magnetization jump in the \mbox{$J$-$Q$-$h$} chain for $q>q_{\rm min}$, but it would be difficult to extract an accurate value for $q_{\rm min}$ from these data alone. 
Instead, we will determine a precise value of $q_{\rm min}$ using the condition in Eq.~(\ref{eq:jumpCondition1}). 
To do this, we first note that the effect of the $Q$ term on the two-magnon subspace is a short-range \textit{attractive} interaction, albeit an unusual one including correlated hopping (see Appendix \ref{s:ajq} for a detailed analysis). 
From Eq.~(\ref{eq:e1}) we know that $\bar E_1 = -2J$ and we can then find a condition on $\bar E_2$ for a bound state to form as a result of this attraction:
\begin{align}
\bar E_2 \leq -4J \label{eq:jqjc1}.
\end{align}

With this in hand, we may interpret the magnetization jumps seen in the QMC data for $q>q_{\rm min}$ as follows: At higher  magnetization densities, this short-range attractive force dominates, causing the gas of magnetic excitations to suddenly condense, producing a magnetization jump. 
Indeed, when the magnetization was fixed at a nonequilibrium value in the QMC calculations (for example, $m=1/2,q=1.2$), we observed phase separation: the chain would separate into a region with magnetization density $m_c$ and another region that was fully polarized. 
Therefore, we may identify $q_{\rm min}$ with the threshold value of $q$ at which Eq.~(\ref{eq:jqjc1}) is first satisfied.

In Fig. \ref{e2} we plot $\bar E_2 (J=1, Q=q)$; we can determine $q_{\rm min}$ by finding the smallest value of $q$ that satisfies Eq.~(\ref{eq:jqjc1}). 
In this way, we obtain $q_{\rm min} = 0. \bar 2 = \frac{2}{9}$ to machine precision for all $L>6$. 
For $q<q_{\rm min}$, finite-size effects result in an \textit{overestimate} of $\bar E_2 (L\rightarrow \infty)$, and for $q>q_{\rm min}$, they result in an \textit{underestimate}. 
At exactly $q=q_{\rm min}$, these effects cancel and $\bar E_2$ becomes independent of $L$ (for $L>6$). 
Note that $q_{\rm min} < q_c$ (the VBS critical point); we should not expect $q_c$ and $q_{\rm min}$ to match since the magnetization jump occurs not from the VBS but from the critical XY state and they are arise from completely different mechanisms.

\begin{figure}
\includegraphics[width=75mm]{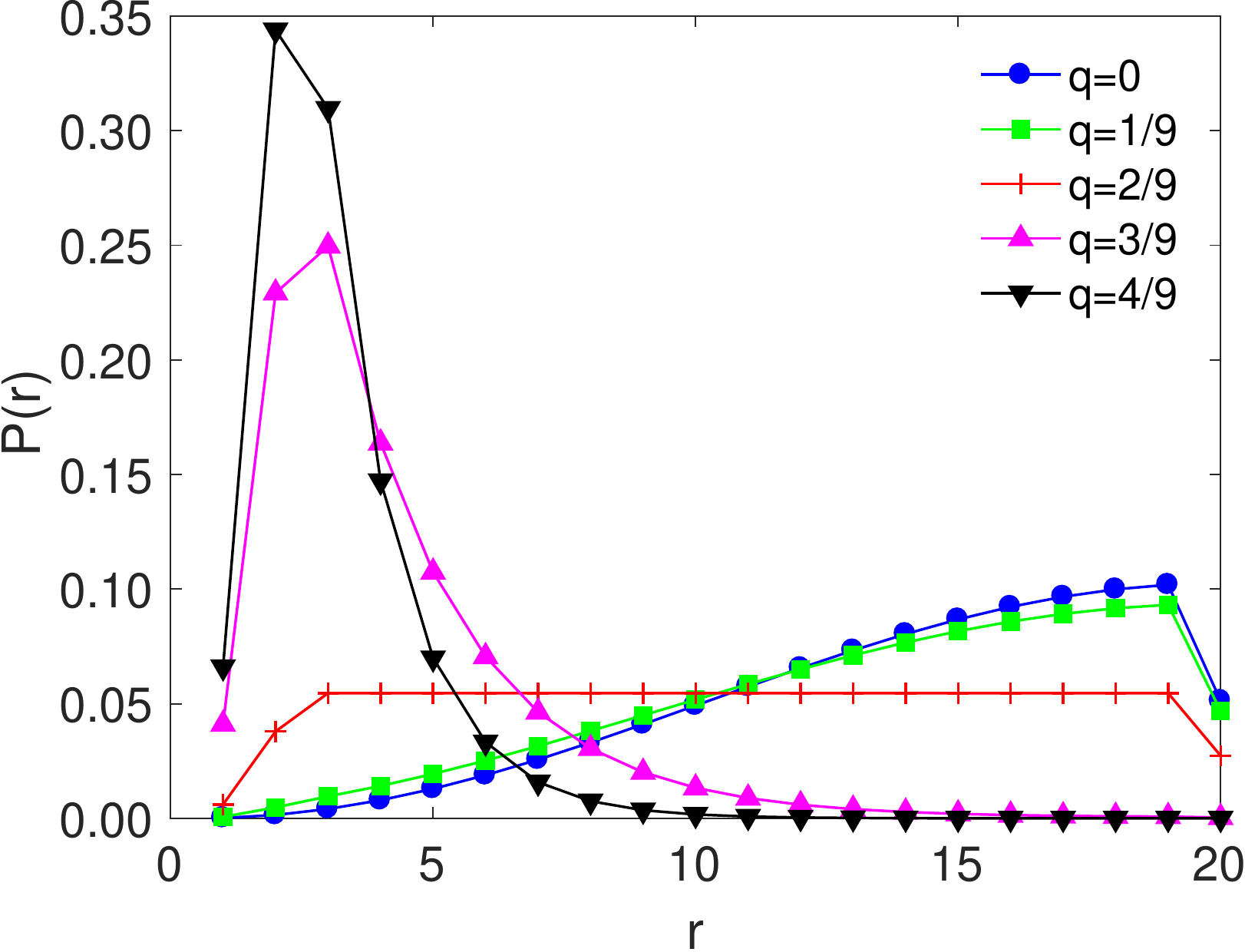}
\caption{The probability $P(r)=\braket{\psi_0(r)|\psi_0 (r)}$  of the particles being separated by distance $r$ in the lowest state in the two-magnon sector ($m_z = S-2$) of the \mbox{$J$-$Q$-$h$} chain.  \label{pr}}
\end{figure}

In Fig. \ref{pr}, we plot the probability density $|\psi_0 (r)|^2$ for $L=40$ chains at several values of $q$ ($r$ is the magnon separation in the separated center-of-mass and relative-coordinate basis as defined in detail in Appendix~\ref{s:ajq}). 
For $q<q_{\rm min}$, the magnons scatter off one another with a finite-range effective repulsive interaction, and the relative wave function takes on (essentially) the form of a particle in a box.
For $q>q_{\rm min}$, the magnons scatter with a finite-range effective \textit{attractive} interaction, in this case the wave function has an exponential decay for $r\geq 3$, indicating a bound state. 
At $q=q_{\rm min}$, magnons cross between these two regimes, scattering off one another acquiring no phase and, thus, their wave function and ground-state energy resemble that of two noninteracting magnons, with $\bar E_2 (J,Q_{\rm min}) = 2 \cdot (-2J)$. 
The wave function is exactly constant in the bulk ($3<r<L/2-1$). 
This completely flat wave function in the bulk at $q_{\rm min}$ (which we will discuss analytically further below) is \textit{not} a generic behavior at the onset of a bound state; typically, one would find an exponentially decaying short-distance disturbance (as we will show in one case of the $J_1$-$J_2$ chain in Sec.~\ref{s:j1j2}). 
As $q \rightarrow q_{\rm min}$ from above, the expectation value of the separation between the magnons diverges. 

Finally, with the precise value of $q_{\rm min}$ determined in this way, we use large-scale QMC data to confirm (Fig. \ref{varq}) that $q_{\rm min}$ is indeed the beginning of an instability that leads to a macroscopic discontinuity in the magnetization. 
This is consistent with previous work, \cite{aligia2000,kecke2007} where bound states of such magnons have been found to be the cause of metamagnetism in spin chains, though previously the attractive interactions were directly related to geometric frustration (which is not present in the \mbox{$J$-$Q$} chain; the $Q$ term competes in a different way against AFM order).

\subsection{An exact solution at $q_{\rm min}$ \label{s:qmin}}

The absence of finite-size effects, the fact that $q_{\rm min}$ is a ratio of small whole numbers, and the flat wave function are remarkable and they provide a hint that there may be an unusually simple analytic solution of the two-magnon system at $q_{\rm min}$. 
Using the separation basis, we can combine Eqs.~(\ref{eq:jsep}) and (\ref{eq:qsep}), set $J=1,Q=q$, and the total momentum $K=0$ and write the Hamiltonian as:
\begin{align}
& -4H= \label{eq:hsep} \\
& \begin{pmatrix}
4+q		&	4+2q		&	q		&	0		&	0		&	\cdots 		&	0\\
4+2q		&	8+4q		&	4+2q	&	0		&	0		&	\cdots		&	0\\
q			&	4+2q		&	8+q	&	4		&	0		&	\cdots		&	0\\
0			&	0			&	4		&	8		&	4		&	0			&	\cdots \\
\vdots		&	\vdots		&	\ddots		&	\ddots	&	\ddots	&	\ddots		&	\ddots	\\
\vdots		&	\vdots		&	0		&	4		&	8		&	4			&	0 \\
\vdots		&	\vdots		&	\ddots	&	0		&	4		&	8			&	4 \sqrt 2 \\
0			&	0			& \cdots	&	\cdots	&	0		&	4\sqrt 2	&	8 
\end{pmatrix}. \nonumber
\end{align} 

Using the simple-looking numerical result for the wave function \mbox{$\psi(r,q=q_{\rm min})$} in Fig.~\ref{pr} as inspiration for finding the ground state, we will now assume (and later confirm) that it has the following form:
\begin{align}
\ket{\psi} \propto & + a \ket{1} + b \ket{2} + c \ket{3} + \sum \limits_{r=4}^{L/2-1}  \ket{r} +  d \ket{L/2}.  \label{eq:psi}
\end{align} 
The wave function is constant in the bulk, but at the edges of the $r$ subspace the state has weights $a,b,c,d$ that can be easily determined. 
Acting on $\ket{\psi}$ with $H$ in Eq.~(\ref{eq:hsep}) produces a set of five equations which can be solved for $a,b,c,d,q_{\rm min}$ and the eigenvalue $\lambda$ with the following results: 
\begin{subequations}
\begin{align}
& a = \frac{1}{3},~~ b = \frac{5}{6},~~ c = 1,~~ d = \frac{1}{\sqrt 2}, \\
& \lambda = -4J, \\
& q_{\rm min} = \frac{2}{9}. 
\end{align}
\end{subequations}
When this solution is plugged back into Eq.~(\ref{eq:psi}), we indeed find an exact match for the numerical results for $q=q_{\rm min}$ plotted in Fig.~\ref{pr}. 

\subsection{Excluded mechanisms for metamagnetism}

We will now discuss some other processes known to cause magnetization jumps, such as localization, \cite{schnack2001,schulenburg2002,richter2004} magnetization plateaus, \cite{honecker2004} and multi-polar phases \cite{balents2015} and then show that none explain the behavior of the \mbox{$J$-$Q$-$h$} chain. 
Although metamagnetism \textit{can} be caused by localization,   \cite{schnack2001,schulenburg2002,richter2004} this cannot be the cause in this case because the \mbox{$J$-$Q$-$h$} chain has no intrinsic disorder and we see no other signs of localization.
Metamagnetism has also been observed in a study of the frustrated FM Heisenberg chain, \cite{sudan2009,arlego2011,balents2015} which has a sequence of multipolar phases. 
If such phases existed near $q_{\rm min}$, we would observe a ``cascade'' of jumps. 
First, the smallest possible jump of $\Delta m_z = 2$ would appear, but then for slightly larger values of $q > q_{\rm min}$, there should be a series of system-size-independent jumps, $\Delta m_z= 3,4,5,...$ until, eventually, a macroscopic jump in the thermodynamic limit. 
Based on exact diagonalization of chains up to $L=28$, we see no evidence of such size-independent jumps in the \mbox{$J$-$Q$-$h$} chain nor do we see any evidence of such an effect in our QMC data. 

\begin{figure}
\includegraphics[width=75mm]{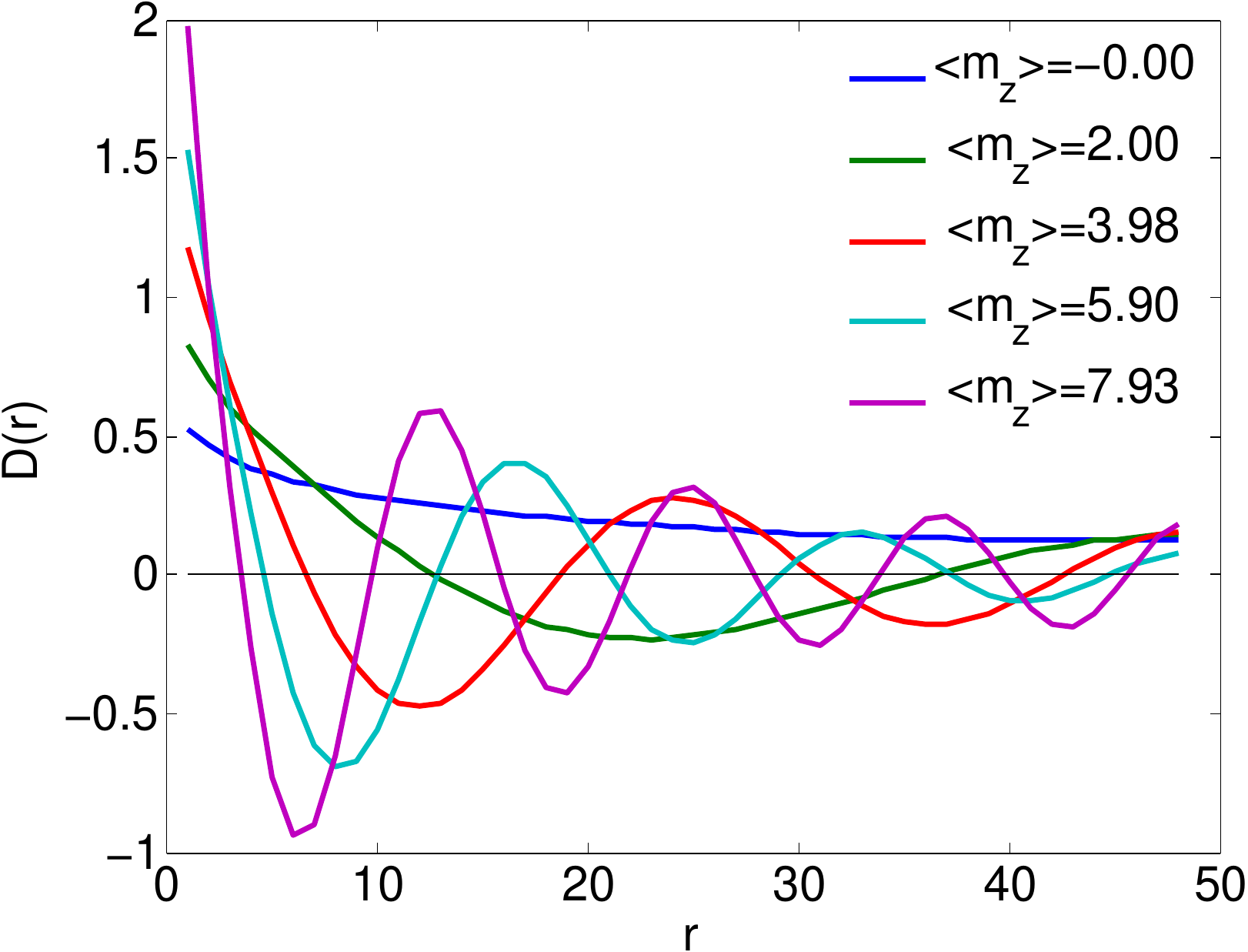}
\caption{Alternating dimer-dimer correlation function, defined in Eq.~(\ref{eq:dr}), for several values of the magnetization in chains of length \mbox{$L=96$} at $\beta=12$, $q=1.2$.   \label{bcor}}
\end{figure}

A jump in the magnetization can also be connected to a magnetization plateau. \cite{honecker2004} 
There is no sign of a magnetization plateau in Figs.~\ref{varq} or \ref{magvh}, but to conclusively rule this out, we can also examine spin correlation functions. 
A magnetization plateau indicates the presence of a gap between different spin states and is allowed (by an extension of the Lieb-Shultz-Mattis theorem) only when the magnetization per unit cell, $m$, obeys the constraint that $(S-m)$ is an integer. \cite{oshikawa1997} 
For a $S=1/2$ chain, this can only occur by breaking translational symmetry. 
We examined the alternating part of the dimer-dimer correlation function, $D(r)$, for signs of translational symmetry breaking. 
This correlation function is defined as
\begin{align}
D(r) = (-1)^r \left[ B(r) - B(r+1) \right] \label{eq:dr},
\end{align}
where $B(r) = \braket{P_{i,i+1} P_{i+r,i+1+r}}$ measures the correlations between bond singlet densities. 
In the VBS-ordered phase, $D(r)$ has the form \mbox{$D(r) \propto (e^{-r/\xi} + D_0)$}, where $D_0$ is the VBS order parameter.  
In Fig.~\ref{bcor}, we plot $D(r)$ for several different values of the magnetization. 
For $m_z>0$, $D(r)$ develops long-wavelength oscillations with a wavelength proportional to the inverse magnetization density $\lambda \propto 1/m$ (a similar effect was predicted in 1D quantum fluids by Haldane\cite{haldane1981}), but we find no evidence of broken symmetry.  
The $S^z$ spin correlations develop a similar pattern of long-wavelength oscillations, and also show no signs of a symmetry-broken state. 
As a final test, we looked at chains with open boundaries and found no signs of symmetry-broken states in that case either. 

\section{Metamagnetism in the \mbox{$J_1$-$J_2$} chain \label{s:j1j2}}

In the \mbox{$J$-$Q$-$h$} chain, the $Q$ term favors AFM ordering at the classical level (where the singlet-projection aspect is not manifested), but nonetheless it produces a short-range \textit{attractive} interaction for low densities of magnons (against a saturated background). 
Other Hamiltonians with these features may exist, and since they also lack frustration, they are likely to be understudied. 
Using the recipe from the \mbox{$J$-$Q$-$h$} chain: (AFM first-neighbor interaction) + (short-range attractive magnon-magnon interaction), a natural challenge is then to create a minimal unfrustrated quantum spin model which also exhibits this effect using only two-spin interactions. We can construct a minimal model by adding an anisotropic \textit{ferro}magnetic (FM) next-nearest-neighbor term to the AFM Heisenberg chain. 
We will now show that a bound state of magnons occurs in the \mbox{$J_1$-$J_2$} model, but only with spin anisotropy in the $J_2$ term, i.e., with the Hamiltonian
\begin{align}
H_{J_1J_2} = & -J_1 \sum \limits_i P_{i,i+1}  \label{eq:j1j2ham} \\
& - J_2 \sum \limits_i \left[ \frac{1}{4} - S^z_i S^z_{i+2} -\frac{\Delta}{2} \left( S^+_i S^-_{i+2} + H.c. \right) \right]. \nonumber
\end{align}
Here, we have defined $\Delta$ in such a way as to guarantee that the $S^z S^z$ interactions of the second-neighbor term are FM for all $J_2<0$. 

When $\Delta = 1$, $J_2>0$ (AFM), Eq.~(\ref{eq:j1j2ham}) becomes the simplest example of a frustrated spin model; this case has been well studied. \cite{majumdar1,majumdar2,tonegawa1987,farnell1993,gerhardt1997,gerhardt1998,aligia2000,kecke2007,sudan2009,arlego2011,kolezhuk2012,soos2016}
Several papers have presented evidence of metamagnetism in the \mbox{$J_1$-$J_2$} chain in this regime for both the isotropic \cite{dmitriev2006,kecke2007,sudan2009,arlego2011,kolezhuk2012} and anisotropic  \cite{gerhardt1998,hirata1999,aligia2000,kolezhuk2012} cases. 
Naively, a FM second-neighbor term is trivial since it does not produce frustration; with an AFM first-neighbor coupling it would serve to strengthen the AFM order. 
Probably for this reason, the FM $J_2$ case has been almost completely overlooked in the literature. 
Only a few papers \cite{farnell1993,gerhardt1997,lu2006} have considered this case and none of them investigated the possibility of metamagnetism. 
Metamagnetism \textit{has} been reported in the 2D and 3D AFM Ising model with a FM second-neighbor term, \cite{landau1972} and a physically equivalent square-lattice-gas model. \cite{rikvold1983}

As with the \mbox{$J$-$Q$-$h$} chain, we will identify the onset of a bound state of two magnons on a fully polarized FM background. 
As we discussed in Sec.~\ref{s:qmin}, such a bound state is a possible signature of metamagnetism, but not a guarantee of it (although in any case the onset of a bound state is an important aspect of other possible transitions). 
We define the criterion for the bound state as
\begin{align}
\bar E_2(j,\Delta) \leq & 2 \bar E_1 (j,\Delta) \label{eq:j1j2jumpcon},
\end{align}
where $J_1=1$ (AFM), $j \equiv -J_2/J_1$ ($j>0$ corresponding to FM $J_2$).
The magnon binding energy is therefore 
\begin{align}
\Xi (j, \Delta) \equiv 2 \bar E_1 -\bar E_2, \label{eq:xidef}
\end{align}
such that $\Xi>0$ indicates the presence of a bound state. 

The exact one-magnon energy, $\bar E_1$, is derived in Appendix~\ref{s:aj1j2} and displayed in Eq.~(\ref{eq:j1j2e1}). 
The two magnon energy, $\bar E_2$, can be determined numerically using the separation-basis Hamiltonian constructed from $H_{J_1}$ and $H_{J_2}$ [Eqs.~(\ref{eq:j1sep}) and (\ref{eq:j2sep})]. 
We will limit ourselves to the unstudied case of FM $J_2$ ($j>0$) and, for simplicity, we will consider only three values of $\Delta$: $\Delta = 1$ (the isotropic case); $\Delta = 0$ (the Ising case); and $\Delta = -1$ (where the Ising interaction is FM and the XY interactions are AFM). 

\begin{figure}
\includegraphics[width=75mm]{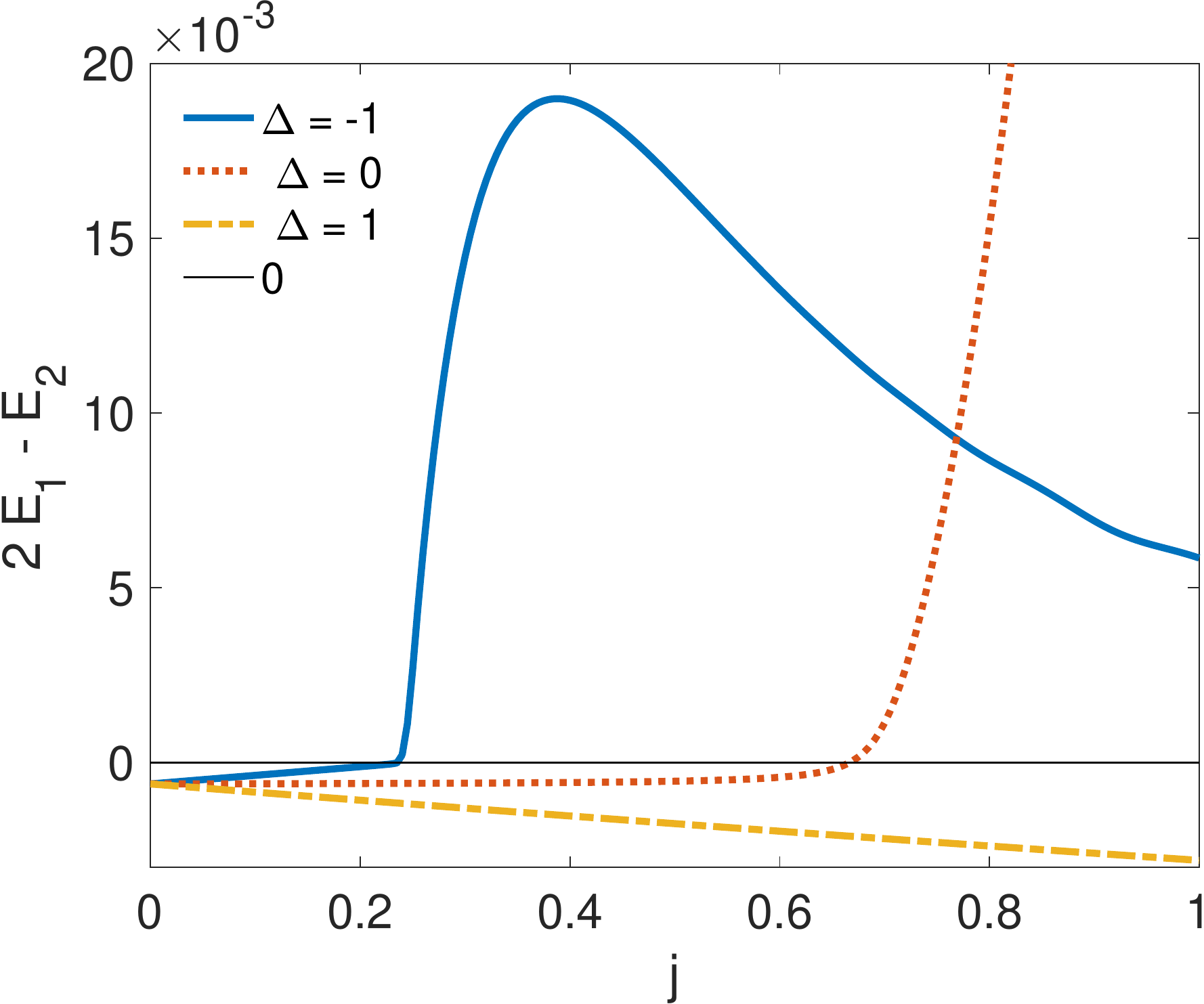}
\caption{The binding energy defined in Eq.~(\ref{eq:xidef}) for a \mbox{$J_1$-$J_2$} chain with $j\equiv -J_2/J_1$ and anisotropy parameters $\Delta = 0, \pm 1$. Here a relatively small system ($L=128$) is used, to make it easier to see the crossings. When $\Xi (j,\Delta) > 0$, there is a bound state of two magnons. \label{j1j2xo}}
\end{figure}

In Fig.~\ref{j1j2xo}, we plot $\Xi (j, \Delta)$ versus $j$ for chains of length \mbox{$L=128$}. 
For large $L$, the level crossing occurs at a very shallow angle and the lines in Fig.~\ref{j1j2xo} tend to overlap; we therefore use a small system size here to make the crossing more clear. 
In the isotropic case, $\Delta = 1$, $\Xi(j,0) <0$ for all $j$ and there is no bound state.  
In the Ising case, $\Delta = 0$, there is a level crossing at $j_{\rm min}=\frac{2}{3}$ (verified to machine precision for chains up to $L=4096$), and for $\Delta = -1$ the bound state occurs above $j_{\rm min} = 0.236067977499$ (to machine precision for $L\geq 32$). 

For $\Delta=0$, the wave function takes on a flat form at $j_{\rm min} = \frac{2}{3}$. 
Using the same approach we used for $q_{\rm min}$ in Sec.~\ref{s:qmin}
\begin{align}
\ket{\psi} \propto &  -\frac{1}{3} \ket{1} + \sum \limits_{r=2}^{L/2-1} (-1)^r \ket{r} +  \frac{1}{\sqrt 2} \ket{L/2}. \label{eq:del0}
\end{align}
Except for the alternating sign, this is almost identical to the flat wave function for the \mbox{$J$-$Q$-$h$} chain at $q_{\rm min}$ and finite-size effects are similarly absent at this point. 
For $\Delta=-1$, the form for the ground state at $j_{\rm min}$ is nearly flat with an exponential tail,
\begin{align}
\ket{\psi} \propto &  \sum \limits_{r=1}^{L/2-1}\hskip-1mm (-1)^r ( 1-a e^{-r/b} ) \ket{r} +  \frac{(-1)^{L/2}}{\sqrt 2} \ket{L/2}, \label{eq:delm1}
\end{align}
where $a=1.447$ and $b=2.078$, based a fit to the numerical wave function (solving directly involves a transcendental equation that we have not studied further). 
In this case, finite-size effects are present, but vanish exponentially in $L$. 
The existence of this two-magnon bound state may be a precursor to a macroscopic magnetization jump, but there is no guarantee that it produces the required instability to multi-magnon bound states. 
Confirming the existence of this transition with large-scale calculations would be an interesting topic for a future study, although the regime \mbox{$\Delta < 0$} is inaccessible to QMC due to the sign problem.

\section{Zero-scale-factor universality\label{s:zsf}}

\begin{figure*}%
\includegraphics[width=160mm]{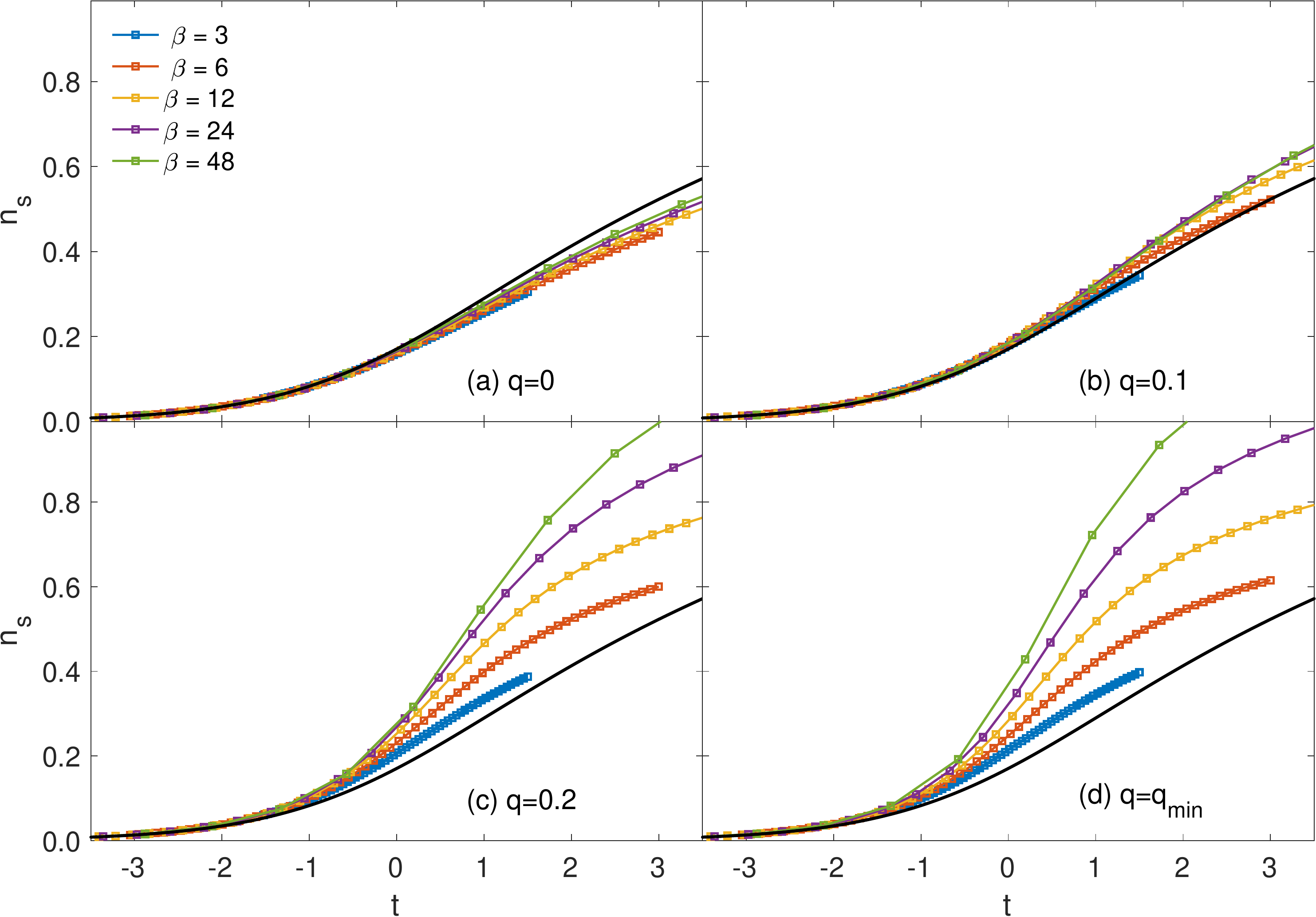}
\caption{Test of zero-factor scaling using the rescaled  density, Eq.~(\ref{eq:ns}) of flipped spins near saturation for a \mbox{$J$-$Q$-$h$} chain of 96 sites for several different inverse temperatures $\beta$ and values of the coupling ratio $q$ (in different panels as indicated). The results are graphed versus the rescaled magnetic field $t \equiv \beta (h_s-h)$. The black lines are the exact predicted universal function, Eq.~(\ref{eq:univ}) with the bare magnon mass $M=1$. \label{fig:zsf}}
\end{figure*}

The critical behavior that has become known as zero-scale-factor universality occurs when response functions are universal functions of bare coupling
constants with no non-universal factors. \cite{sachdev1994} 
Zero-scale-factor universality is expected to apply in one-dimensional systems whenever there is a continuous quantum phase transition that corresponds to the smooth onset of a nonzero ground state expectation value for a conserved density variable. 
In spin chains, the most well-studied realization is the field-tuned transition from the Haldane-gapped singlet state of integer spin chains to a state in which one polarization of triplet magnons ($S=1$ quasiparticle excitations above the singlet state) condenses to give a nonzero magnetization density. 

The saturation transition in the \mbox{$J$-$Q$-$h$} chain provides a different realization: the magnons are now single spin-flip excitations above the saturated (i.e., fully polarized) ground state (the same magnons as in Sec.~\ref{s:qmin}), and the transition in question is the transition from the saturated state to the partially polarized critical state. 
When this transition is continuous the density of magnons turns on continuously.
Moreover, the density of these magnons is conserved by virtue of the U(1) symmetry of spin rotations about the $z$ axis.
Therefore, the magnetization density, Eq.~(\ref{eq:magden}), in the vicinity of the saturation transition, is expected to obey the following form [from Eq.~(1.23) of Ref.~\onlinecite{sachdev1994}]: 
\begin{align}
\braket{m} = g \mu_B \left( \frac{2 M}{\hbar^2 \beta} \right)^{1/2} \mathcal{M} (\mu \beta), \label{eq:e123}
\end{align}
where $M$ is the magnon mass and \mbox{$\mu =(h_s - h)$}. 

The single magnon dispersion (\ref{eq:e1}) obeys the low-energy quadratic form $\epsilon(k) \propto k^2/(2M)$, with $M=1$ (in our units where $J=1$) independently of $Q$. 
The $Q$ term gives rise to an additional contribution to the hopping if two magnons are within three lattice spacings of each other.
Considering the low magnon density and repulsive magnon-magnon interactions, we only expect a negligible renormalization of $M$ due to this correlated hopping term. 
We define \mbox{$\braket{m} = g \mu_B \braket{n}$}, where $n$ is the density of flipped spins and \mbox{$\mu =(h_s - h)$}. 
In this way, the field above the saturation value represents the ``gap'' for these magnetic excitations and a negative $\mu$ corresponds to $h>h_s$. 
We insert these definitions into Eq.~(\ref{eq:e123}): 
\begin{align}
\braket{n} \left( \frac{\hbar^2 \beta}{2M} \right)^{1/2} =  \mathcal{M} [\beta (h_s -h)]
\end{align}
To simplify further, we set $\hbar =1$ and define the rescaled field $t \equiv \beta (h_s -h)$:
\begin{align}
n_s (q, t) \equiv \braket{n} \sqrt{\frac{\beta}{2M}} =  \mathcal{M}(t) \label{eq:ns}
\end{align}
We will henceforth call $n_s$ the rescaled magnon density. 
The one-dimensional case is unique here, in that there is a known analytic form \cite{sachdev1994} for the universal scaling function $\mathcal{M}(t)$:
\begin{align}
\mathcal{M} (t) = & \frac{1}{\pi} \int \limits_0^\infty dy \frac{1}{e^{y^2-t}+1} = -\frac{1}{2 \sqrt \pi} \mathrm{Li}_{1/2} (-e^{t}) \label{eq:univ}
\end{align}
In the limit $|t| \rightarrow \infty$, the polylogarithm simplifies and the universal function becomes
\begin{align}
\mathcal{M}(t) =  &
\begin{cases} 
      \frac{\sqrt{t}}{\pi} 		& t \rightarrow \infty ,\\
      \frac{e^t}{2 \sqrt \pi} 	& t \rightarrow -\infty ,
\end{cases}
\end{align}
but we will use the full form without approximations.

The critical behavior of the magnetization near the saturation field was recently studied using the finite-temperature Bethe ansatz in the case of the standard $S=1/2$ Heisenberg chain, \cite{he2017} and detailed comparisons were also made with experimental results for AFM chain \cite{kono2015,jeong2015} and ladder \cite{watson2001} systems. 
In order to explicitly test the validity of the zero-scale-factor universality, we here analyze our data in a different manner from Ref.~\onlinecite{he2017}.

\begin{figure}%
\includegraphics[width=80mm]{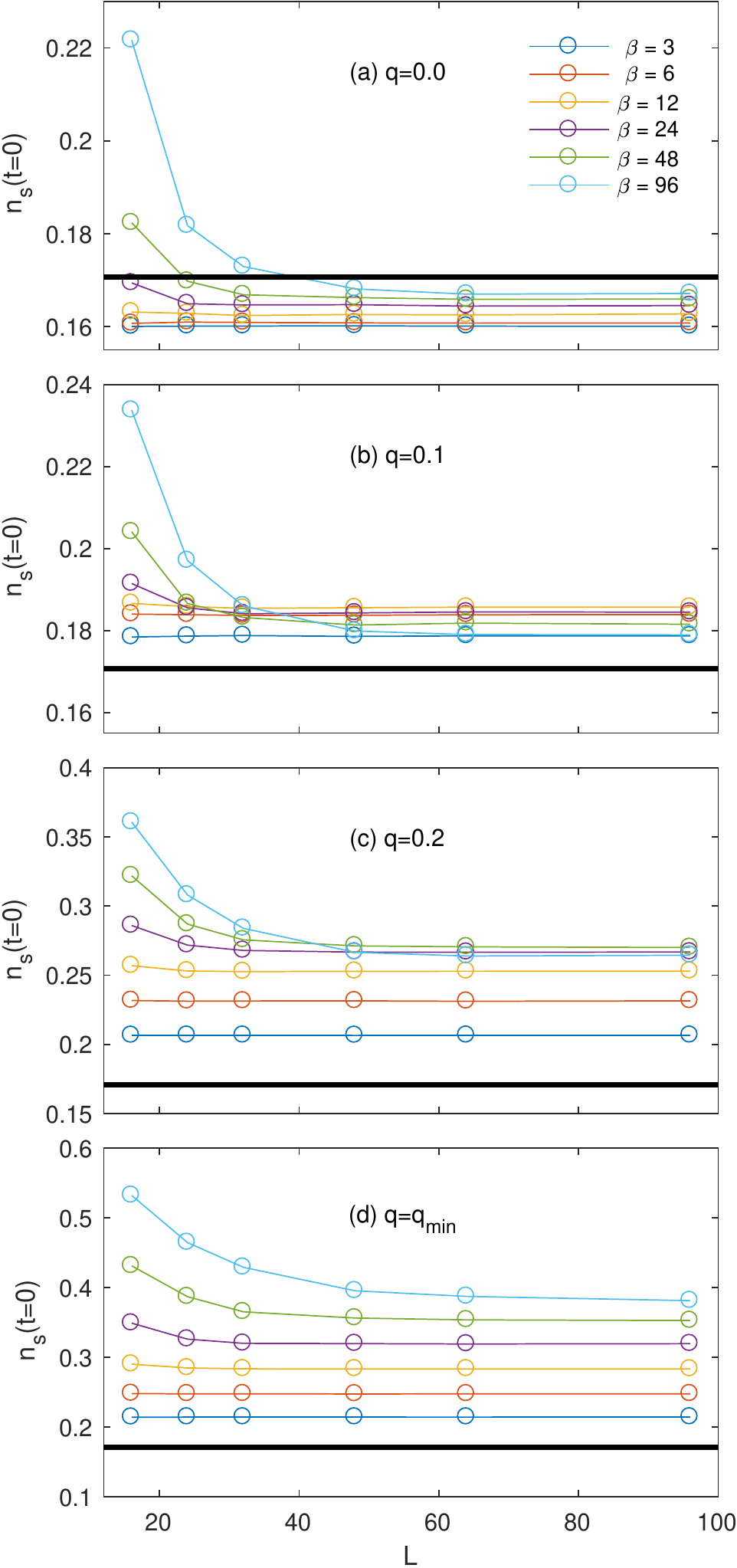}
\caption{Finite-size behavior of the zero-factor scaled magnon density, Eq.~(\ref{eq:ns}), for the \mbox{$J$-$Q$-$h$} chain at \mbox{$t\equiv \beta(h_s-h)=0$} for several different inverse temperatures $\beta$ and values of the coupling ratio $q$ (in different panels as indicated). In all cases, the error bars are smaller than the markers. The black horizontal lines in each panel show the value from the exact universal function, Eq.~(\ref{eq:univ}), with the bare magnon mass $M=1$. \label{fig:zsffss} }
\end{figure}

In Fig.~\ref{fig:zsf}, we plot the rescaled density, $n_s$, as a function of the rescaled field, $t$, for $L=96$ \mbox{$J$-$Q$-$h$} chains near the saturation transition for \mbox{$q=0, \, 0.1, \, 0.2$} and \mbox{$q= q_{\rm min}$}. 
In all these cases, $h_s = 2J$ (see Fig.~\ref{hcrit}). 
The rescaled data collapse reasonably well for $q=0$, as shown in Fig.~\ref{fig:zsf}(a), although it is also clear that we have not quite reached the asymptotic large-$\beta$ scaling limit (the curves for even the highest $\beta$ values still exhibit some drift). 
We have investigated other system sizes to ensure that finite-size corrections are not important here (see also Fig.~\ref{fig:zsffss}).
Figs. \ref{fig:zsf}(b)--\ref{fig:zsf}(d), we apply the same rescaling and find that the agreement with the theory becomes progressively worse for increasing $q$. 
The curves for different temperatures still collapse rather well onto one another for $t<0$, but the collapsed data no longer match the shape of the universal function, even if we choose $M$ different from the bare value $M=1$ in the single-magnon dispersion (and, as already noted, we do not expect any significant renormalization of $M$ due to many-body effects at these low magnon densities). 
Additionally, the quality of the collapse itself deteriorates for $t>0$. 
As expected, for $q>q_{\rm min}$ (not shown) the zero-factor scaling fails completely: the magnons now interact attractively, and there is discontinuity in $\braket{n}$ which cannot be rescaled to match an analytic function. 

\begin{figure}%
\includegraphics[width=80mm]{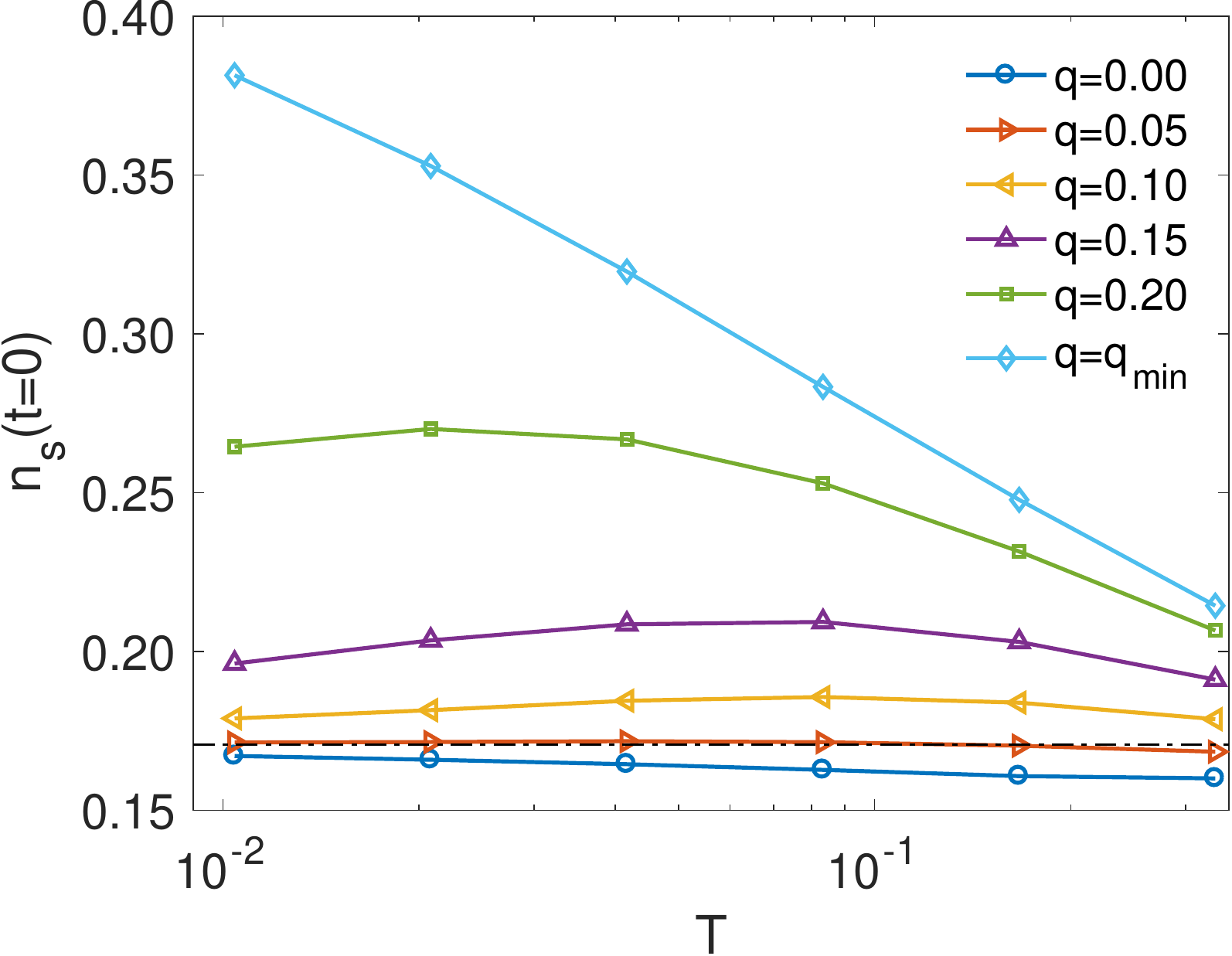}
\caption{Temperature dependence of the rescaled magnon density, Eq.~(\ref{eq:ns}), for an $L=96$ \mbox{$J$-$Q$-$h$} chain at $h=h_s$ and several values of the coupling ratio $q$. Error bars are smaller than the markers. The black dashed line shows the exact asymptotic ($T\to 0$) value from the universal function, Eq.~(\ref{eq:univ}), setting the bare magnon mass $M=1$. \label{fig:zsftemp}}
\end{figure}

It is not obvious from Fig.~\ref{fig:zsf} that this scaling form works at all for $q \neq 0$. 
To explore this more carefully, we examine the finite-size scaling of $n_s$ with the field set to saturation \mbox{($t=0$)} in Fig.~\ref{fig:zsffss}. 
In this case, the exact universal function has no dependence on $\beta$, but in all panels of Fig.~\ref{fig:zsffss}, there remains significant $\beta$ dependence. 
Clearly, we have not yet reached the low temperatures (high $\beta$) where the universal form applies without significant corrections (as seen in Fig.~\ref{fig:zsftemp}, exceedingly low temperatures are required to observe this convergence, especially for $q>0$). 
The $\beta$ dependence becomes stronger for larger values of $q$.
We also see non-monotonic $\beta$-dependence for $q=0.1$ and $0.2$, which manifests as the crossing of lines in Fig.~\ref{fig:zsffss}(b)-(c). 
This non-monotonic behavior explains why, in Fig.~\ref{fig:zsf}(b)-(c), the agreement with the exact function sometimes gets \textit{worse} for increasing $\beta$. 
At \mbox{$q= q_{\rm min}$} the agreement with the exact form is far worse and $n_s$ at $t=0$ shows no signs of convergence. 
Instead, it shows a monotonic increase with $\beta$; this supports the notion that $q_{\rm min}$ is a tricritical point with a different scaling behavior. 
The cross-overs seen in the $\beta$-dependence for \mbox{$0 < q < q_{\rm min}$} should then be due to a cross-over temperature related to the tricritical point.

We take a closer look at the temperature dependence in Fig.~\ref{fig:zsftemp}, where we plot $n_s$ at $t=0$ versus the temperature $T=\beta^{-1}$ for a fixed size $L=96$. 
Here, the cross-over behavior is clear and we know from Fig.~\ref{fig:zsffss} that finite-size effects are not important at this size. 
The dashed black line represents the exact value of the universal function from Eq.~(\ref{eq:univ}) evaluated at $t=0$, $M=1$.
For $q=0$, we can see that the results converge monotonically toward the expected value from below. 
With $q=0.05$, $n_s(t=0)$ is extremely close to the exact value, but a careful examination shows that the behavior is non-monotonic with a broad maximum before a flattening out at lower temperatures, consistent with asymptotic convergence to the expected value. 
For $q=0.1$, the behavior of $n_s(t=0)$ is similar and more clearly visible on the scale of Fig.~\ref{fig:zsftemp}. 
For $q=0.15, \,0.2$, there is a maximum at lower $T$ but we cannot see the convergence to the universal value when $T\to 0$, although we expect this to take place at still lower temperatures. 
For $q=q_{\rm min}$, the behavior is essentially a logarithmic divergence, but we do not know the power of the logarithm. 
All these behaviors are consistent with a low-energy description with a $|\psi|^4$-type field theory, where the coefficient of the $|\psi|^4$ term vanishes at $q_{\rm min}$, and at this point the critical behavior is controlled not by the zero-scale-factor theory but by the marginal $|\psi|^6$ term (causing the logarithmic scaling). 
The cross-over temperature between the two critical behaviors, as manifested by the maximum in $n_s(t=0)$ versus $T$, should gradually approach \mbox{$T=0$} as \mbox{$q \to q_{\rm min}$} from below, as we indeed observe in Fig.~\ref{fig:zsftemp}.

We summarize our findings on the zero-scale-factor universality. 
In Fig.~\ref{fig:zsf}, we observe that this scaling works very well for $q=0$, but the scaling appears to work poorly for $0 < q \leq q_{\rm min}$. 
By examining finite-size scaling of the rescaled magnetization in Fig.~\ref{fig:zsffss}, we observe non-monotonic temperature dependence for $0 < q < q_{\rm min}$. 
Finally, in Fig.~\ref{fig:zsftemp}, we plot $n_s$ as a function of $T$ for $t=0$, here we can see that for all $q<q_{\rm min}$, $n_s$ appears to converge toward the exact value at $T\rightarrow 0$. 
As $q$ approaches $q_{\rm min}$, the temperature required to observe convergence becomes extremely low due to the influence of the tricriticality. 
These results are consistent with the behavior predicted by the theory: the zero-scale-factor universality applies for all $q<q_{\rm min}$ and fails only at the tricritical point $q_{\rm min}$. 
Finally, this divergence occurs for $q_{\rm min}=\frac{2}{9}$ which confirms the results of the level-crossing analysis documented in Sec.~\ref{s:qmin}.

\section{Conclusions and Discussion \label{s:discussion}}

In this paper, we have studied the \mbox{$J$-$Q$} chain in the presence of an external magnetic field using range of techniques including exact diagonalization, a few-magnon expansion, and a parallelized quantum replica exchange within the SSE QMC method.  
We have established the existence of a metamagnetic transition (i.e., magnetization jump) to the saturated state for $q \geq q_{\rm min} = \frac{2}{9}$, a first-order quantum phase transition caused by the onset of a bound state of magnons (flipped spins on a FM background). 
This proves that metamagnetism can occur in the absence of both frustration \textit{and} intrinsic anisotropy. 
The magnetization jump begins with zero magnitude at $q=q_{\rm min}$ and increases gradually in magnitude with $q$. 
Below $q_{\rm min}$, magnons interact with a finite-range effectively repulsive interaction. 
Above $q_{\rm min}$, magnons interact with a finite-range effectively \textit{attractive} interaction, despite the absence of any explicitly FM interactions. 
At the onset of the jump, magnons become noninteracting (for sufficiently low density) and the problem of two magnons in a polarized background can be solved analytically. 
The point at which two magnons bind represents the onset of an instability where an arbitrary number of magnons attract to form a macroscopic magnetization jump. 
Motivated by the work presented here, the existence of metamagnetism in the \mbox{$J$-$Q$-$h$} chain and our proposed mechanism for it have been confirmed by calculations using the density matrix renormalization group. \cite{mao2017}

It may be difficult to find an experimental realization of the \mbox{$J$-$Q$} model itself, but interactions similar to the $Q$ term can appear in effective models of spin-phonon chains where the phonons have been integrated out. \cite{suwa2014} 
Thus, spin-phonon systems may possibly harbor metamagnetism even in the absence of longer-range frustrated Heisenberg exchange interactions. 
We again stress that $q_{\rm min}$, the threshold for metamagnetism, is significantly smaller than $q_c$, the threshold for dimerization; therefore, spin-phonon systems may also harbor metamagnetism even if the spin-phonon coupling is insufficiently strong to produce dimerization. \cite{sandvik1999}

The saturation transition in the \mbox{$J$-$Q$-$h$} chain is rich, and we have shown that the magnetization near saturation obeys a zero-scale-factor universality \cite{sachdev1994} at $q=0$, which becomes increasingly difficult to observe as $q$ is increased above about $ \approx 0.1$. 
This is explained by the influence of the tricritical point at $q_{\rm min}$, where the low-energy effective field theory changes, leading to a different criticality and cross-over behavior. 
The most natural scenario is that the coefficient of $|\psi|^4$ vanishes in the $|\psi|^4$ effective field theory for the saturation transition at the threshold for formation of the two-magnon bound state, thereby allowing the $|\psi|^6$ term to control the scaling behavior of the saturation transition at this threshold. 
This term is marginal in spatial dimension $d=1$ since the dynamical exponent for the transition is $z=2$, implying the presence of logarithmic violations of scaling at $q=q_{\rm min}$. 
In our QMC data, we indeed observe logarithmic scaling of the magnetization density exactly at $q_{\rm min}$.

Using the same two-magnon approach from the \mbox{$J$-$Q$-$h$} chain, we have studied the \mbox{AFM-FM} \mbox{$J_1$-$J_2$} chain with anisotropy $\Delta$ in the $J_2$ term [see Eq.~(\ref{eq:j1j2ham})]. 
We have found that for $\Delta=0,-1$, there is a bound state of magnons for $j>j_{\rm min}$ with $j_{\rm min} = \frac{2}{3},\, 0.236$ respectively. 
It is likely that these bound states will cause a magnetization jump to saturation in this model, but we have not investigated this possibility using large-scale simulations.  
The $S^z$ interactions in the $J_2$ term are in both cases FM and have the effect of reinforcing the zero-field ground state correlations. 
Thus, they produce no frustration in the conventional sense, but still lead to nontrivial behavior. 
To our knowledge, no study has previously attempted to find metamagnetism in the AFM-FM \mbox{$J_1$-$J_2$} chain, and this would be an excellent topic for a future study using the density matrix renormalization group method, which is well suited for frustrated one-dimensional systems. 
Such a study could also confirm whether the zero-scale-factor universality is obeyed by the \mbox{$J_1$-$J_2$} chain near saturation and compare the breakdown as $j\rightarrow j_{\rm min}$ to the breakdown that occurs in the \mbox{$J$-$Q$-$h$} chain. 
Indeed, the \mbox{AFM-FM} \mbox{$J_1$-$J_2$} chain may be generally understudied due to its lack of conventional frustration. 
The existence of a nontrivial behavior in this previously overlooked unfrustrated spin chain may mean that there are other phenomena to explore in such naively trivial Hamiltonians. 

The methods developed for this work, including the parallelized replica exchange quantum Monte Carlo program, are now being extended to study the 2D \mbox{$J$-$Q$-$h$} model in the presence of a magnetic field. 
Our preliminary results indicate magnetization jumps above a coupling ratio $q_{\rm min}$ and a similar mechanism of bound states of magnons as in one dimension. 
In two dimensions we do not expect zero-scale-factor universality close to saturation for $q < q_{\rm min}$, because we are then at the upper critical dimension (2+2) of this theory. 
Logarithmic corrections may then be expected for all $q < q_{\rm min}$, and the behavior at $q_{\rm min}$ is unclear at present.

The lower metamagnetic bound, $q_{\rm min}$ is less than $q_c$ (the dimerization transition), and indeed, the physics of metamagnetism appears completely unrelated to the physics of the dimerization transition. 
More generally, we note the utility of $J$-$Q$-type models for studies of phenomena normally associated with frustration due to competing exchange interactions, e.g., $J_1$-$J_2$ Heisenberg models. 
Due to the absence of sign problems, these models can be studied with QMC simulations in any number of dimensions, while techniques for frustrated models (e.g., the density matrix renormalization group technique) are restricted to one dimension and (still) relatively small two-dimensional systems. 
VBS physics, in particular the AFM--VBS transition, has so far been the primary goal of studies with \mbox{$J$-$Q$} models, and our present work now adds metamagnetism and high-field scaling to this repertoire of phenomena accessible to QMC simulations of this family of ``designer Hamiltonians''. \cite{kaul2013}

\section*{Acknowledgements}
The work of A.I. and A.W.S. was supported by the NSF under Grant No.~DMR-1410126. 
A.I. acknowledges support from the APS-IUSSTF Physics PhD Student Visitation Program for a visit to K.D. at the Tata Institute of Fundamental Research in Mumbai. The computational work reported in this paper was performed in part on the Shared Computing Cluster administered by Boston University's Research Computing Services. 
We gratefully acknowledge the help of H.-G. Luo, B.B. Mao and C. Cheng who helped us discover an error in the \mbox{$J_1$-$J_2$} calculation in our original preprint. 

\appendix

\section{Few magnons in the \mbox{$J$-$Q$-$h$} chain \label{s:ajq}}

Continuing from Sec. \ref{s:qmin}, we will attempt to  find $q_{\rm min}$, the value of $q$ where the jump first appears. 
To do this, we will look for a direct level crossing between saturated state $m_z = S$ and the state with two flipped spins $m_z=S-2$ and therefore we must calculate $E(m_z,J,Q,L)$ for $m_z=S,S-1,S-2$.  
Finding energy of the saturated state is trivial: there are no places for a singlet projection operator to act, so $H\ket{m_z = S} = -hS$. 
If we add a single \mbox{spin-down} site (magnon), the Heisenberg term produces a tight-binding-like effective Hamiltonian on this flipped spin: the diagonal terms give it an on-site energy and the off-diagonal terms allow it to hop to neighboring sites. 
A $Q$ term cannot act on this single-magnon state.
The one-magnon state is a one-body problem with the analytic solution: 
\begin{align}
E_1 =& -J (1-\cos k) -h(S-1) \label{eq:b4rot}
\end{align}
for periodic boundary conditions. 

For purposes of algorithmic convenience, we will perform a `sublattice rotation,' a unitary transformation on one sublattice which rotates $S^+_j \rightarrow S^-_j$. 
This transformation has the effect of flipping the signs of all off-diagonal terms in the Hamiltonian without changing the spectrum. \cite{sandvik2011computational}
After the sublattice rotation, Eq.~(\ref{eq:b4rot}) becomes:
\begin{align}
E_1 =& -J (1+\cos k) -h(S-1) \label{eq:e1}
\end{align}
Note that the sign of the $\cos k$ term has changed.
With $J>0$, the ground state has momentum $k=0$; therefore
\begin{align}
E_1 = & -2J-h(S-1)
\end{align}
for all $L$. 
For $q<q_{\rm min}$, the saturation field is determined by a direct level crossing between $E_0$ and $E_1$, so the saturation field is independent of $Q$:
\begin{align}
h_s(q<q_{\rm min}) = & 2J \label{eq:hsatj}
\end{align}
 
For the two-magnon case, we can begin in the basis of the positions of each flipped spin: $\ket{x_1,x_2}$; the size of this basis is $L(L-1)/2$. 
We will assume that $L$ is even. 
We can reduce this two-particle problem to single particle problem using translation invariance. 
Consider a basis of the center-of-mass position and the distance between the spin-down sites: $\ket{X, r}$. 
The center of mass takes on the values $X\equiv x_2+x_1=3, 4, 5, 6,... (2L-1)$ and the separation takes on the values $r\equiv \min \left[ x_2-x_1, L+x_1-x_2 \right]=1,2,...L/2$. 
The Hamiltonian is translation-invariant for the center-of-mass coordinate, $X$, so we can consider momentum states: $\ket{K,r}$. 
Where $K$ is the center-of-mass momentum and $r$ is the separation between the magnons. 
\begin{align}
K_n = \frac{2\pi n}{L}, \,\,\, n=0,1,2,...L-1 \label{eq:com}
\end{align}
For a given $K_n$, \mbox{$r=1,2,3,...r_{\rm max}$}.  
We must be careful with our definitions to avoid double counting states. 
For even-$n$, \mbox{$r=1,2,...L/2$}, but for odd-$n$, \mbox{$r=1,2,...L/2-1$}. 
Thus, for each of the $L/2$ even-$n$ momentum states, there are $L/2$ $r$-states, and for each of the $L/2$ odd-$n$ momentum states, there are $L/2-1$ $r$-states, for a total of $L(L-1)/2$ states. 

Now consider how the Heisenberg term acts on a two-magnon state $\ket{x_1,x_2}$:
\begin{align}
H_J & \ket{x_1,x_2}  =  -2J \ket{x_1,x_2} -  \frac{J}{2} \Big[ \ket{x_1+1,x_2} \nonumber \\
 + & \ket{x_1-1,x_2} + \ket{x_1,x_2+1} + \ket{x_1,x_2-1} \Big].
\end{align}
There are two ways to hop the magnons toward each other, two ways to hop them away from each other, and four ways to leave them where they are, each with magnitude $-J/2$. 
In the separation basis, this becomes: 
\begin{align}
H_J \ket{r>2} = &-2J \ket{r} -\frac{J}{2}(1+e^{-iK})\ket{r-1} \nonumber \\
& -\frac{J}{2}(1+e^{iK})\ket{r+1}
\end{align}
Thus, in the `bulk' ($1<r<L/2$), the result is very similar to the one-magnon problem. 
For $r=1$, there are two slight modifications: the spin-down sites are hardcore bosons (they cannot hop across each other) and the diagonal term is only $-J$. 
For $r=\frac{L}{2}-1$ and $\frac{L}{2}$, there are slight modifications due to the boundary conditions. 
Put this all together and we get:
\begin{align}
& H_J  = -J \times \label{eq:jsep} \\
& \begin{pmatrix} 
1 					& \frac{1+e^{iK}}{2} 	& 0 			& \ldots 		&			&	\\ 
\frac{1+e^{-iK}}{2} & 2 					& \frac{1+e^{iK}}{2}	& 0				& \ldots 	&	\\
0				& \frac{1+e^{-iK}}{2}		& 2				& \frac{1+e^{iK}}{2}	& 0			&	\ldots	\\
\vdots 	&	\vdots	&	\ddots	&	\ddots		&	\ddots	& 			&	&	\\
\vdots 	&	\vdots	&				&\ddots	&	\ddots		&	\ddots	& 			&	\\
& &	\frac{1+e^{-iK}}{2}	& 2				& \frac{1+e^{iK}}{2}			& \underline{0}		 \\
& & 0				& \frac{1+e^{-iK}}{2}	& 2		 				& \underline{\frac{1+e^{iK}}{\sqrt 2}} 		\\
& &  			& \underline{0}	& \underline{\frac{1+e^{-iK}}{\sqrt 2}}	&	\underline{2}		\\ 
\end{pmatrix} \nonumber
\end{align}
where the last row and last column (\underline{underlined} entries) are omitted in the odd-$n$ momentum sectors. 

Now consider the $Q$ term, which only contributes for $r\leq3$, so we can represent it as a $3\times3$ matrix:
\begin{align}
H_Q = -\frac{Q}{4}
\begin{pmatrix}
1			& 1+e^{iK}		& e^{iK}	\\
1+e^{-iK}	& 2(1+\cos K)	& 1+e^{iK}	\\
e^{-iK}		& 1+e^{-iK}		& 1	\\
\end{pmatrix} \label{eq:qsep}
\end{align}
Somewhat counterintuitively, the $Q$ term produces an effective \textit{attractive} interaction by lowering the energy of states where the flipped spins are separated by no more than three lattice spacings. 
This will be the key to producing the magnetization jump.

Now, we have the energies of each magnetization sector:
\begin{subequations}
\begin{align}
E_S = & -hS,	\\
E_1 = & \bar E_1(J,Q,L) -h(S-1), \\
E_2 = & \bar E_2 (J,Q,L) -h(S-2),
\end{align} \label{eq:energies}
\end{subequations}
where $\bar E_n$ is the ground state energy of the zero-field $n$-magnon chain. 
In order to find $q_{\rm min}$, we must first find the saturation field $h_s$ by demanding that $E_S=E_2$:
\begin{align}
h_s = -\frac{1}{2} \bar E_2 (J,Q,L). \label{eq:hc}
\end{align}
To guarantee a direct level crossing between $m_z = S-2$ and $m_z = S$, require $E_1 \geq E_S = E_2$:
\begin{align}
-h_s S \leq & \bar E_1 -h_s (S-1),	\\
h_s \geq & -\bar E_1. \label{eq:hc2}
\end{align}
Combining Eqs.~(\ref{eq:hc}) and (\ref{eq:hc2}) and eliminating $h_s$, we find a condition for $q_{\rm min}$:
\begin{align}
\bar E_2 \leq 2 \bar E_1.  \label{eq:jumpCondition}
\end{align}
This condition is also essentially the condition for an attractive interaction: the energy for two magnons is \textit{less} than twice the single-magnon energy because the interactions \textit{lower} the total energy. 
From Eq.~(\ref{eq:e1}), we know that $\bar E_1 = -2J$, so we can find a condition on $\bar E_2$ for the existence of a jump:
\begin{align}
\bar E_2 \leq -4J. \label{eq:jqjc}
\end{align}

\section{Derivation of magnetization jump in \mbox{$J_1$-$J_2$} chain \label{s:aj1j2}}

The anisotropic $J_2$ term is given by
\begin{align}
H_{J_2} = & - J_2 \times \label{eq:j2ham} \\
&\sum \limits_i \left[ \frac{1}{4} - S^z_i S^z_{i+2} -\frac{\Delta}{2} \left( S^+_i S^-_{i+2} + H.c. \right) \right]. \nonumber 
\end{align}
We will set $J_2=-j$ ($j>0$ is \textit{ferro}magnetic) and follow the same steps from Appendix \ref{s:ajq}. 
First, we need the one-magnon energy, which can be derived in much the same way we derived the one-magnon energy for the \mbox{$J$-$Q$-$h$} chain: 
\begin{align}
\bar E_1 (j, \Delta) = & -J_1 ( 1- \cos k) - J_2 (1- \Delta \cos 2k), \\
\bar E_1 (j, \Delta) = & -1 + \cos k + j- j\Delta \cos 2k. \label{eq:e1j1j2}
\end{align}
Note that here we do \textit{not} use the sublattice rotation employed in Appendix \ref{s:ajq}; this difference can be seen by comparing Eq.~(\ref{eq:e1j1j2}), where the potential energy ($-1$) and kinetic energy ($\cos k$) terms have the \textit{opposite} sign, to Eq.~(\ref{eq:e1}), where they have the \textit{same} sign. 
For $\Delta \geq 0$, $\bar E_1$ is always minimized by $k=\pi$. 
For $\Delta < 0$, $k_{\rm min}$ can take on two values
\begin{align}
k_{\rm min}(j,\Delta) =  &
\begin{cases} 
      \pi, 										& (j\Delta) \geq -1/4 \\
      \arccos \left( \frac{1}{4j\Delta}\right), 	& (j \Delta) < -1/4 .
\end{cases}
\end{align}
This means that the ground state energy for one magnon is given by:
\begin{align}
\bar E_1 (j,\Delta) =  &
\begin{cases} 
      -2 + j (1-\Delta)							& (j\Delta) \geq -1/4 \\
      -1 + j (1+\Delta) + \frac{1}{8j\Delta}	& (j \Delta) < -1/4 
\end{cases} \label{eq:j1j2e1}
\end{align}

Now we want to write the two-magnon Hamiltonian in the separation basis (as defined in Appendix \ref{s:ajq}). 
We have already worked out the separation basis Hamiltonian for the $J_1$ term in Eq.~(\ref{eq:jsep}), but in this case we cannot use the sublattice rotation. 
Reversing the sublattice rotation done to Eq.~(\ref{eq:jsep}), we arrive at a form for $H_{J_1}$:
\begin{align}
& H_{J_1}  = J_1 \times \label{eq:j1sep} \\
& \begin{pmatrix} 
-1 					& \frac{1+e^{iK}}{2} 	& 0 			& \ldots 		&			&	\\ 
\frac{1+e^{-iK}}{2} & -2 					& \frac{1+e^{iK}}{2}	& 0				& \ldots 	&	\\
0				& \frac{1+e^{-iK}}{2}		& -2			& \frac{1+e^{iK}}{2}	& 0			&	\ldots	\\
\vdots 	&	\vdots	&	\ddots	&	\ddots		&	\ddots	& 			&	&	\\
\vdots 	&	\vdots	&				&\ddots	&	\ddots		&	\ddots	& 			&	\\
& &	\frac{1+e^{-iK}}{2}	& -2				& \frac{1+e^{iK}}{2}			& \underline{0}		 \\
& & 0				& \frac{1+e^{-iK}}{2}	& -2		 				& \underline{\frac{1+e^{iK}}{\sqrt 2}} 		\\
& &  			& \underline{0}	& \underline{\frac{1+e^{-iK}}{\sqrt 2}}	&	\underline{-2}		\\ 
\end{pmatrix} \nonumber
\end{align}
Notice that Eq.~(\ref{eq:j1sep}) is identical to Eq.~(\ref{eq:jsep}), except for the signs of the off-diagonal terms. 
$H_{J_2}$ can be derived in the same way that we derived the separation basis Hamiltonian for the Heisenberg chain, Eq.~(\ref{eq:jsep}). 
Applying the same logic to the $J_2$ term, we arrive at: 
\begin{widetext}
\begin{align}
H_{J_2} (K)= & \label{eq:j2sep}\\
j&\begin{pmatrix}
2-\Delta\cos K					& 0	& -\frac{\Delta(1+e^{2iK})}{2}	& 0								& 0								& \cdots 	&	\\
0								& 1	& 0								& -\frac{\Delta(1+e^{2iK})}{2}	& 0								& \cdots	&	\\
-\frac{\Delta(1+e^{-2iK})}{2}	& 0	& 2								& 0								& -\frac{\Delta(1+e^{2iK})}{2}	& \cdots	&	\\
								& \ddots & \ddots					& \ddots						& \ddots 						& \ddots 	& 	\\
& -\frac{\Delta(1+e^{-2iK})}{2}	& 0	& 2								& 0	& -\frac{\Delta(1+e^{2iK})}{2} 	& \underline{0}	\\
& 0								& -\frac{\Delta(1+e^{-2iK})}{2}& 0	& 2	& 0								& \underline{-\frac{\Delta(1+e^{2iK})}{\sqrt 2}} \\
& 0			& 0					& -\frac{\Delta(1+e^{-2iK})}{2} & 0	& 2-\Delta\cos K				&  \underline{0}  \\
& \underline{0} & \underline{0}	& \underline{0}				& \underline{-\frac{\Delta(1+e^{-2iK})}{\sqrt 2}}		&  \underline{0}	& \underline{2}
\end{pmatrix} \nonumber
\end{align} \end{widetext}
where the rows and columns represent $r=1,2,3,...L/2$. 
As in Appendix \ref{s:ajq}, for even-$n$ momentum sectors, $r=1,2,3,... L/2$ and for odd-$n$ momentum sectors the basis is truncated $r=1,2,3,...L/2-1$, so we must cut off the last row and column of Eqs.~(\ref{eq:j1sep}) and (\ref{eq:j2sep}) (the \underline{underlined} entries). 
This approach is based on one used by Kecke et. al. to study the FM-AFM \mbox{$J_1$-$J_2$} chain. \cite{kecke2007}

\bibliography{bibstuff}

\begin{thebibliography}{57}%
\makeatletter
\providecommand \@ifxundefined [1]{%
 \@ifx{#1\undefined}
}%
\providecommand \@ifnum [1]{%
 \ifnum #1\expandafter \@firstoftwo
 \else \expandafter \@secondoftwo
 \fi
}%
\providecommand \@ifx [1]{%
 \ifx #1\expandafter \@firstoftwo
 \else \expandafter \@secondoftwo
 \fi
}%
\providecommand \natexlab [1]{#1}%
\providecommand \enquote  [1]{``#1''}%
\providecommand \bibnamefont  [1]{#1}%
\providecommand \bibfnamefont [1]{#1}%
\providecommand \citenamefont [1]{#1}%
\providecommand \href@noop [0]{\@secondoftwo}%
\providecommand \href [0]{\begingroup \@sanitize@url \@href}%
\providecommand \@href[1]{\@@startlink{#1}\@@href}%
\providecommand \@@href[1]{\endgroup#1\@@endlink}%
\providecommand \@sanitize@url [0]{\catcode `\\12\catcode `\$12\catcode
  `\&12\catcode `\#12\catcode `\^12\catcode `\_12\catcode `\%12\relax}%
\providecommand \@@startlink[1]{}%
\providecommand \@@endlink[0]{}%
\providecommand \url  [0]{\begingroup\@sanitize@url \@url }%
\providecommand \@url [1]{\endgroup\@href {#1}{\urlprefix }}%
\providecommand \urlprefix  [0]{URL }%
\providecommand \Eprint [0]{\href }%
\providecommand \doibase [0]{http://dx.doi.org/}%
\providecommand \selectlanguage [0]{\@gobble}%
\providecommand \bibinfo  [0]{\@secondoftwo}%
\providecommand \bibfield  [0]{\@secondoftwo}%
\providecommand \translation [1]{[#1]}%
\providecommand \BibitemOpen [0]{}%
\providecommand \bibitemStop [0]{}%
\providecommand \bibitemNoStop [0]{.\EOS\space}%
\providecommand \EOS [0]{\spacefactor3000\relax}%
\providecommand \BibitemShut  [1]{\csname bibitem#1\endcsname}%
\let\auto@bib@innerbib\@empty
\bibitem [{\citenamefont {Sandvik}(2007)}]{sandvik2007}%
  \BibitemOpen
  \bibfield  {author} {\bibinfo {author} {\bibfnamefont {A.~W.}\ \bibnamefont
  {Sandvik}},\ }\href {\doibase 10.1103/PhysRevLett.98.227202} {\bibfield
  {journal} {\bibinfo  {journal} {Phys. Rev. Lett.}\ }\textbf {\bibinfo
  {volume} {98}},\ \bibinfo {pages} {227202} (\bibinfo {year}
  {2007})}\BibitemShut {NoStop}%
\bibitem [{\citenamefont
  {Sandvik}(2010{\natexlab{a}})}]{sandvik2011computational}%
  \BibitemOpen
  \bibfield  {author} {\bibinfo {author} {\bibfnamefont {A.~W.}\ \bibnamefont
  {Sandvik}},\ }\href {\doibase 10.1063/1.3518900} {\bibfield  {journal}
  {\bibinfo  {journal} {AIP Conf. Proc.}\ }\textbf {\bibinfo {volume} {1297}},\
  \bibinfo {pages} {135} (\bibinfo {year} {2010}{\natexlab{a}})}\BibitemShut
  {NoStop}%
\bibitem [{\citenamefont {Kaul}\ \emph {et~al.}(2013)\citenamefont {Kaul},
  \citenamefont {Melko},\ and\ \citenamefont {Sandvik}}]{kaul2013}%
  \BibitemOpen
  \bibfield  {author} {\bibinfo {author} {\bibfnamefont {R.~K.}\ \bibnamefont
  {Kaul}}, \bibinfo {author} {\bibfnamefont {R.~G.}\ \bibnamefont {Melko}}, \
  and\ \bibinfo {author} {\bibfnamefont {A.~W.}\ \bibnamefont {Sandvik}},\
  }\href {\doibase 10.1146/annurev-conmatphys-030212-184215} {\bibfield
  {journal} {\bibinfo  {journal} {Annu. Rev. Condens. Matter Phys.}\ }\textbf
  {\bibinfo {volume} {4}},\ \bibinfo {pages} {179} (\bibinfo {year}
  {2013})}\BibitemShut {NoStop}%
\bibitem [{\citenamefont {Sanyal}\ \emph {et~al.}(2011)\citenamefont {Sanyal},
  \citenamefont {Banerjee},\ and\ \citenamefont {Damle}}]{sanyal2011}%
  \BibitemOpen
  \bibfield  {author} {\bibinfo {author} {\bibfnamefont {S.}~\bibnamefont
  {Sanyal}}, \bibinfo {author} {\bibfnamefont {A.}~\bibnamefont {Banerjee}}, \
  and\ \bibinfo {author} {\bibfnamefont {K.}~\bibnamefont {Damle}},\
  }\href@noop {} {\bibfield  {journal} {\bibinfo  {journal} {Phys. Rev. B}\
  }\textbf {\bibinfo {volume} {84}},\ \bibinfo {pages} {235129} (\bibinfo
  {year} {2011})}\BibitemShut {NoStop}%
\bibitem [{\citenamefont {Tang}\ and\ \citenamefont
  {Sandvik}(2011)}]{tang2011a}%
  \BibitemOpen
  \bibfield  {author} {\bibinfo {author} {\bibfnamefont {Y.}~\bibnamefont
  {Tang}}\ and\ \bibinfo {author} {\bibfnamefont {A.~W.}\ \bibnamefont
  {Sandvik}},\ }\href {\doibase 10.1103/PhysRevLett.107.157201} {\bibfield
  {journal} {\bibinfo  {journal} {Phys. Rev. Lett.}\ }\textbf {\bibinfo
  {volume} {107}},\ \bibinfo {pages} {157201} (\bibinfo {year}
  {2011})}\BibitemShut {NoStop}%
\bibitem [{\citenamefont {Tang}\ and\ \citenamefont
  {Sandvik}(2015)}]{tang2014}%
  \BibitemOpen
  \bibfield  {author} {\bibinfo {author} {\bibfnamefont {Y.}~\bibnamefont
  {Tang}}\ and\ \bibinfo {author} {\bibfnamefont {A.~W.}\ \bibnamefont
  {Sandvik}},\ }\href {\doibase 10.1103/PhysRevB.92.184425} {\bibfield
  {journal} {\bibinfo  {journal} {Phys. Rev. B}\ }\textbf {\bibinfo {volume}
  {92}},\ \bibinfo {pages} {184425} (\bibinfo {year} {2015})}\BibitemShut
  {NoStop}%
\bibitem [{\citenamefont {Lou}\ \emph {et~al.}(2009)\citenamefont {Lou},
  \citenamefont {Sandvik},\ and\ \citenamefont {Kawashima}}]{lou2009}%
  \BibitemOpen
  \bibfield  {author} {\bibinfo {author} {\bibfnamefont {J.}~\bibnamefont
  {Lou}}, \bibinfo {author} {\bibfnamefont {A.~W.}\ \bibnamefont {Sandvik}}, \
  and\ \bibinfo {author} {\bibfnamefont {N.}~\bibnamefont {Kawashima}},\ }\href
  {\doibase 10.1103/PhysRevB.80.180414} {\bibfield  {journal} {\bibinfo
  {journal} {Phys. Rev. B}\ }\textbf {\bibinfo {volume} {80}},\ \bibinfo
  {pages} {180414} (\bibinfo {year} {2009})}\BibitemShut {NoStop}%
\bibitem [{\citenamefont {Sandvik}(2010{\natexlab{b}})}]{sandvik2010}%
  \BibitemOpen
  \bibfield  {author} {\bibinfo {author} {\bibfnamefont {A.~W.}\ \bibnamefont
  {Sandvik}},\ }\href {\doibase 10.1103/PhysRevLett.104.177201} {\bibfield
  {journal} {\bibinfo  {journal} {Phys. Rev. Lett.}\ }\textbf {\bibinfo
  {volume} {104}},\ \bibinfo {pages} {177201} (\bibinfo {year}
  {2010}{\natexlab{b}})}\BibitemShut {NoStop}%
\bibitem [{\citenamefont {Jin}\ and\ \citenamefont {Sandvik}(2013)}]{jin2013}%
  \BibitemOpen
  \bibfield  {author} {\bibinfo {author} {\bibfnamefont {S.}~\bibnamefont
  {Jin}}\ and\ \bibinfo {author} {\bibfnamefont {A.~W.}\ \bibnamefont
  {Sandvik}},\ }\href {\doibase 10.1103/PhysRevB.87.180404} {\bibfield
  {journal} {\bibinfo  {journal} {Phys. Rev. B}\ }\textbf {\bibinfo {volume}
  {87}},\ \bibinfo {pages} {180404} (\bibinfo {year} {2013})}\BibitemShut
  {NoStop}%
\bibitem [{\citenamefont {Tang}\ and\ \citenamefont
  {Sandvik}(2013)}]{tang2013}%
  \BibitemOpen
  \bibfield  {author} {\bibinfo {author} {\bibfnamefont {Y.}~\bibnamefont
  {Tang}}\ and\ \bibinfo {author} {\bibfnamefont {A.~W.}\ \bibnamefont
  {Sandvik}},\ }\href {\doibase 10.1103/PhysRevLett.110.217213} {\bibfield
  {journal} {\bibinfo  {journal} {Phys. Rev. Lett.}\ }\textbf {\bibinfo
  {volume} {110}},\ \bibinfo {pages} {217213} (\bibinfo {year}
  {2013})}\BibitemShut {NoStop}%
\bibitem [{\citenamefont {Sandvik}\ and\ \citenamefont
  {Kurkij\"arvi}(1991)}]{sandvik1991}%
  \BibitemOpen
  \bibfield  {author} {\bibinfo {author} {\bibfnamefont {A.~W.}\ \bibnamefont
  {Sandvik}}\ and\ \bibinfo {author} {\bibfnamefont {J.}~\bibnamefont
  {Kurkij\"arvi}},\ }\href {\doibase 10.1103/PhysRevB.43.5950} {\bibfield
  {journal} {\bibinfo  {journal} {Phys. Rev. B}\ }\textbf {\bibinfo {volume}
  {43}},\ \bibinfo {pages} {5950} (\bibinfo {year} {1991})}\BibitemShut
  {NoStop}%
\bibitem [{\citenamefont {Sylju\aa{}sen}\ and\ \citenamefont
  {Sandvik}(2002)}]{sandvik_dl}%
  \BibitemOpen
  \bibfield  {author} {\bibinfo {author} {\bibfnamefont {O.~F.}\ \bibnamefont
  {Sylju\aa{}sen}}\ and\ \bibinfo {author} {\bibfnamefont {A.~W.}\ \bibnamefont
  {Sandvik}},\ }\href {\doibase 10.1103/PhysRevE.66.046701} {\bibfield
  {journal} {\bibinfo  {journal} {Phys. Rev. E}\ }\textbf {\bibinfo {volume}
  {66}},\ \bibinfo {pages} {046701} (\bibinfo {year} {2002})}\BibitemShut
  {NoStop}%
\bibitem [{\citenamefont {Hukushima}\ and\ \citenamefont
  {Nemoto}(1996)}]{hukushima1996}%
  \BibitemOpen
  \bibfield  {author} {\bibinfo {author} {\bibfnamefont {K.}~\bibnamefont
  {Hukushima}}\ and\ \bibinfo {author} {\bibfnamefont {K.}~\bibnamefont
  {Nemoto}},\ }\href {\doibase 10.1143/JPSJ.65.1604} {\bibfield  {journal}
  {\bibinfo  {journal} {J. Phys. Soc. Jpn.}\ }\textbf {\bibinfo {volume}
  {65}},\ \bibinfo {pages} {1604} (\bibinfo {year} {1996})}\BibitemShut
  {NoStop}%
\bibitem [{\citenamefont {Sengupta}\ \emph {et~al.}(2002)\citenamefont
  {Sengupta}, \citenamefont {Sandvik},\ and\ \citenamefont
  {Campbell}}]{sengupta2002}%
  \BibitemOpen
  \bibfield  {author} {\bibinfo {author} {\bibfnamefont {P.}~\bibnamefont
  {Sengupta}}, \bibinfo {author} {\bibfnamefont {A.~W.}\ \bibnamefont
  {Sandvik}}, \ and\ \bibinfo {author} {\bibfnamefont {D.~K.}\ \bibnamefont
  {Campbell}},\ }\href {\doibase 10.1103/PhysRevB.65.155113} {\bibfield
  {journal} {\bibinfo  {journal} {Phys. Rev. B}\ }\textbf {\bibinfo {volume}
  {65}},\ \bibinfo {pages} {155113} (\bibinfo {year} {2002})}\BibitemShut
  {NoStop}%
\bibitem [{\citenamefont {Jacobs}\ and\ \citenamefont
  {Lawrence}(1967)}]{jacobs1967}%
  \BibitemOpen
  \bibfield  {author} {\bibinfo {author} {\bibfnamefont {I.~S.}\ \bibnamefont
  {Jacobs}}\ and\ \bibinfo {author} {\bibfnamefont {P.~E.}\ \bibnamefont
  {Lawrence}},\ }\href {\doibase 10.1103/PhysRev.164.866} {\bibfield  {journal}
  {\bibinfo  {journal} {Phys. Rev.}\ }\textbf {\bibinfo {volume} {164}},\
  \bibinfo {pages} {866} (\bibinfo {year} {1967})}\BibitemShut {NoStop}%
\bibitem [{\citenamefont {Stryjewski}\ and\ \citenamefont
  {Giordano}(1977)}]{stryjewski1977}%
  \BibitemOpen
  \bibfield  {author} {\bibinfo {author} {\bibfnamefont {E.}~\bibnamefont
  {Stryjewski}}\ and\ \bibinfo {author} {\bibfnamefont {N.}~\bibnamefont
  {Giordano}},\ }\href {\doibase 10.1080/00018737700101433} {\bibfield
  {journal} {\bibinfo  {journal} {Adv. Phys.}\ }\textbf {\bibinfo {volume}
  {26}},\ \bibinfo {pages} {487} (\bibinfo {year} {1977})}\BibitemShut
  {NoStop}%
\bibitem [{\citenamefont {Aligia}(2000)}]{aligia2000}%
  \BibitemOpen
  \bibfield  {author} {\bibinfo {author} {\bibfnamefont {A.~A.}\ \bibnamefont
  {Aligia}},\ }\href {\doibase 10.1103/PhysRevB.63.014402} {\bibfield
  {journal} {\bibinfo  {journal} {Phys. Rev. B}\ }\textbf {\bibinfo {volume}
  {63}},\ \bibinfo {pages} {014402} (\bibinfo {year} {2000})}\BibitemShut
  {NoStop}%
\bibitem [{\citenamefont {Arlego}\ \emph {et~al.}(2011)\citenamefont {Arlego},
  \citenamefont {Heidrich-Meisner}, \citenamefont {Honecker}, \citenamefont
  {Rossini},\ and\ \citenamefont {Vekua}}]{arlego2011}%
  \BibitemOpen
  \bibfield  {author} {\bibinfo {author} {\bibfnamefont {M.}~\bibnamefont
  {Arlego}}, \bibinfo {author} {\bibfnamefont {F.}~\bibnamefont
  {Heidrich-Meisner}}, \bibinfo {author} {\bibfnamefont {A.}~\bibnamefont
  {Honecker}}, \bibinfo {author} {\bibfnamefont {G.}~\bibnamefont {Rossini}}, \
  and\ \bibinfo {author} {\bibfnamefont {T.}~\bibnamefont {Vekua}},\ }\href
  {\doibase 10.1103/PhysRevB.84.224409} {\bibfield  {journal} {\bibinfo
  {journal} {Phys. Rev. B}\ }\textbf {\bibinfo {volume} {84}},\ \bibinfo
  {pages} {224409} (\bibinfo {year} {2011})}\BibitemShut {NoStop}%
\bibitem [{\citenamefont {Gerhardt}\ \emph {et~al.}(1998)\citenamefont
  {Gerhardt}, \citenamefont {M\"utter},\ and\ \citenamefont
  {Kr\"oger}}]{gerhardt1998}%
  \BibitemOpen
  \bibfield  {author} {\bibinfo {author} {\bibfnamefont {C.}~\bibnamefont
  {Gerhardt}}, \bibinfo {author} {\bibfnamefont {K.-H.}\ \bibnamefont
  {M\"utter}}, \ and\ \bibinfo {author} {\bibfnamefont {H.}~\bibnamefont
  {Kr\"oger}},\ }\href {\doibase 10.1103/PhysRevB.57.11504} {\bibfield
  {journal} {\bibinfo  {journal} {Phys. Rev. B}\ }\textbf {\bibinfo {volume}
  {57}},\ \bibinfo {pages} {11504} (\bibinfo {year} {1998})}\BibitemShut
  {NoStop}%
\bibitem [{\citenamefont {{Hirata}}()}]{hirata1999}%
  \BibitemOpen
  \bibfield  {author} {\bibinfo {author} {\bibfnamefont {S.}~\bibnamefont
  {{Hirata}}},\ }\href@noop {} {\ }\Eprint
  {http://arxiv.org/abs/arXiv:cond-mat/9912066} {arXiv:cond-mat/9912066}
  \BibitemShut {NoStop}%
\bibitem [{\citenamefont {Dmitriev}\ and\ \citenamefont
  {Krivnov}(2006)}]{dmitriev2006}%
  \BibitemOpen
  \bibfield  {author} {\bibinfo {author} {\bibfnamefont {D.~V.}\ \bibnamefont
  {Dmitriev}}\ and\ \bibinfo {author} {\bibfnamefont {V.~Y.}\ \bibnamefont
  {Krivnov}},\ }\href {\doibase 10.1103/PhysRevB.73.024402} {\bibfield
  {journal} {\bibinfo  {journal} {Phys. Rev. B}\ }\textbf {\bibinfo {volume}
  {73}},\ \bibinfo {pages} {024402} (\bibinfo {year} {2006})}\BibitemShut
  {NoStop}%
\bibitem [{\citenamefont {Kecke}\ \emph {et~al.}(2007)\citenamefont {Kecke},
  \citenamefont {Momoi},\ and\ \citenamefont {Furusaki}}]{kecke2007}%
  \BibitemOpen
  \bibfield  {author} {\bibinfo {author} {\bibfnamefont {L.}~\bibnamefont
  {Kecke}}, \bibinfo {author} {\bibfnamefont {T.}~\bibnamefont {Momoi}}, \ and\
  \bibinfo {author} {\bibfnamefont {A.}~\bibnamefont {Furusaki}},\ }\href
  {\doibase 10.1103/PhysRevB.76.060407} {\bibfield  {journal} {\bibinfo
  {journal} {Phys. Rev. B}\ }\textbf {\bibinfo {volume} {76}},\ \bibinfo
  {pages} {060407} (\bibinfo {year} {2007})}\BibitemShut {NoStop}%
\bibitem [{\citenamefont {Sudan}\ \emph {et~al.}(2009)\citenamefont {Sudan},
  \citenamefont {L\"uscher},\ and\ \citenamefont {L\"auchli}}]{sudan2009}%
  \BibitemOpen
  \bibfield  {author} {\bibinfo {author} {\bibfnamefont {J.}~\bibnamefont
  {Sudan}}, \bibinfo {author} {\bibfnamefont {A.}~\bibnamefont {L\"uscher}}, \
  and\ \bibinfo {author} {\bibfnamefont {A.~M.}\ \bibnamefont {L\"auchli}},\
  }\href {\doibase 10.1103/PhysRevB.80.140402} {\bibfield  {journal} {\bibinfo
  {journal} {Phys. Rev. B}\ }\textbf {\bibinfo {volume} {80}},\ \bibinfo
  {pages} {140402} (\bibinfo {year} {2009})}\BibitemShut {NoStop}%
\bibitem [{\citenamefont {Kolezhuk}\ \emph {et~al.}(2012)\citenamefont
  {Kolezhuk}, \citenamefont {Heidrich-Meisner}, \citenamefont {Greschner},\
  and\ \citenamefont {Vekua}}]{kolezhuk2012}%
  \BibitemOpen
  \bibfield  {author} {\bibinfo {author} {\bibfnamefont {A.~K.}\ \bibnamefont
  {Kolezhuk}}, \bibinfo {author} {\bibfnamefont {F.}~\bibnamefont
  {Heidrich-Meisner}}, \bibinfo {author} {\bibfnamefont {S.}~\bibnamefont
  {Greschner}}, \ and\ \bibinfo {author} {\bibfnamefont {T.}~\bibnamefont
  {Vekua}},\ }\href {\doibase 10.1103/PhysRevB.85.064420} {\bibfield  {journal}
  {\bibinfo  {journal} {Phys. Rev. B}\ }\textbf {\bibinfo {volume} {85}},\
  \bibinfo {pages} {064420} (\bibinfo {year} {2012})}\BibitemShut {NoStop}%
\bibitem [{\citenamefont {Landau}(1972)}]{landau1972}%
  \BibitemOpen
  \bibfield  {author} {\bibinfo {author} {\bibfnamefont {D.~P.}\ \bibnamefont
  {Landau}},\ }\href {\doibase 10.1103/PhysRevLett.28.449} {\bibfield
  {journal} {\bibinfo  {journal} {Phys. Rev. Lett.}\ }\textbf {\bibinfo
  {volume} {28}},\ \bibinfo {pages} {449} (\bibinfo {year} {1972})}\BibitemShut
  {NoStop}%
\bibitem [{\citenamefont {Rikvold}\ \emph {et~al.}(1983)\citenamefont
  {Rikvold}, \citenamefont {Kinzel}, \citenamefont {Gunton},\ and\
  \citenamefont {Kaski}}]{rikvold1983}%
  \BibitemOpen
  \bibfield  {author} {\bibinfo {author} {\bibfnamefont {P.~A.}\ \bibnamefont
  {Rikvold}}, \bibinfo {author} {\bibfnamefont {W.}~\bibnamefont {Kinzel}},
  \bibinfo {author} {\bibfnamefont {J.~D.}\ \bibnamefont {Gunton}}, \ and\
  \bibinfo {author} {\bibfnamefont {K.}~\bibnamefont {Kaski}},\ }\href
  {\doibase 10.1103/PhysRevB.28.2686} {\bibfield  {journal} {\bibinfo
  {journal} {Phys. Rev. B}\ }\textbf {\bibinfo {volume} {28}},\ \bibinfo
  {pages} {2686} (\bibinfo {year} {1983})}\BibitemShut {NoStop}%
\bibitem [{\citenamefont {Huerga}\ \emph {et~al.}(2014)\citenamefont {Huerga},
  \citenamefont {Dukelsky}, \citenamefont {Laflorencie},\ and\ \citenamefont
  {Ortiz}}]{huerga2014}%
  \BibitemOpen
  \bibfield  {author} {\bibinfo {author} {\bibfnamefont {D.}~\bibnamefont
  {Huerga}}, \bibinfo {author} {\bibfnamefont {J.}~\bibnamefont {Dukelsky}},
  \bibinfo {author} {\bibfnamefont {N.}~\bibnamefont {Laflorencie}}, \ and\
  \bibinfo {author} {\bibfnamefont {G.}~\bibnamefont {Ortiz}},\ }\href
  {\doibase 10.1103/PhysRevB.89.094401} {\bibfield  {journal} {\bibinfo
  {journal} {Phys. Rev. B}\ }\textbf {\bibinfo {volume} {89}},\ \bibinfo
  {pages} {094401} (\bibinfo {year} {2014})}\BibitemShut {NoStop}%
\bibitem [{\citenamefont {Majumdar}\ and\ \citenamefont
  {Ghosh}(1969{\natexlab{a}})}]{majumdar1}%
  \BibitemOpen
  \bibfield  {author} {\bibinfo {author} {\bibfnamefont {C.~K.}\ \bibnamefont
  {Majumdar}}\ and\ \bibinfo {author} {\bibfnamefont {D.~K.}\ \bibnamefont
  {Ghosh}},\ }\href {\doibase http://dx.doi.org/10.1063/1.1664978} {\bibfield
  {journal} {\bibinfo  {journal} {J. Math. Phys.}\ }\textbf {\bibinfo {volume}
  {10}},\ \bibinfo {pages} {1388} (\bibinfo {year}
  {1969}{\natexlab{a}})}\BibitemShut {NoStop}%
\bibitem [{\citenamefont {Majumdar}\ and\ \citenamefont
  {Ghosh}(1969{\natexlab{b}})}]{majumdar2}%
  \BibitemOpen
  \bibfield  {author} {\bibinfo {author} {\bibfnamefont {C.~K.}\ \bibnamefont
  {Majumdar}}\ and\ \bibinfo {author} {\bibfnamefont {D.~K.}\ \bibnamefont
  {Ghosh}},\ }\href {\doibase http://dx.doi.org/10.1063/1.1664979} {\bibfield
  {journal} {\bibinfo  {journal} {J. Math. Phys.}\ }\textbf {\bibinfo {volume}
  {10}},\ \bibinfo {pages} {1399} (\bibinfo {year}
  {1969}{\natexlab{b}})}\BibitemShut {NoStop}%
\bibitem [{\citenamefont {Soos}\ \emph {et~al.}(2016)\citenamefont {Soos},
  \citenamefont {Parvej},\ and\ \citenamefont {Kumar}}]{soos2016}%
  \BibitemOpen
  \bibfield  {author} {\bibinfo {author} {\bibfnamefont {Z.~G.}\ \bibnamefont
  {Soos}}, \bibinfo {author} {\bibfnamefont {A.}~\bibnamefont {Parvej}}, \ and\
  \bibinfo {author} {\bibfnamefont {M.}~\bibnamefont {Kumar}},\ }\href
  {http://stacks.iop.org/0953-8984/28/i=17/a=175603} {\bibfield  {journal}
  {\bibinfo  {journal} {J. Phys.: Condens. Matter}\ }\textbf {\bibinfo {volume}
  {28}},\ \bibinfo {pages} {175603} (\bibinfo {year} {2016})}\BibitemShut
  {NoStop}%
\bibitem [{\citenamefont {Sachdev}\ \emph {et~al.}(1994)\citenamefont
  {Sachdev}, \citenamefont {Senthil},\ and\ \citenamefont
  {Shankar}}]{sachdev1994}%
  \BibitemOpen
  \bibfield  {author} {\bibinfo {author} {\bibfnamefont {S.}~\bibnamefont
  {Sachdev}}, \bibinfo {author} {\bibfnamefont {T.}~\bibnamefont {Senthil}}, \
  and\ \bibinfo {author} {\bibfnamefont {R.}~\bibnamefont {Shankar}},\ }\href
  {\doibase 10.1103/PhysRevB.50.258} {\bibfield  {journal} {\bibinfo  {journal}
  {Phys. Rev. B}\ }\textbf {\bibinfo {volume} {50}},\ \bibinfo {pages} {258}
  (\bibinfo {year} {1994})}\BibitemShut {NoStop}%
\bibitem [{\citenamefont {Iaizzi}\ and\ \citenamefont
  {Sandvik}(2015)}]{iaizzi2015}%
  \BibitemOpen
  \bibfield  {author} {\bibinfo {author} {\bibfnamefont {A.}~\bibnamefont
  {Iaizzi}}\ and\ \bibinfo {author} {\bibfnamefont {A.~W.}\ \bibnamefont
  {Sandvik}},\ }\href {http://stacks.iop.org/1742-6596/640/i=1/a=012043}
  {\bibfield  {journal} {\bibinfo  {journal} {J. Phys. Conf. Ser.}\ }\textbf
  {\bibinfo {volume} {640}},\ \bibinfo {pages} {012043} (\bibinfo {year}
  {2015})}\BibitemShut {NoStop}%
\bibitem [{\citenamefont {{Sulejmanpasic}}\ \emph {et~al.}()\citenamefont
  {{Sulejmanpasic}}, \citenamefont {{Shao}}, \citenamefont {{Sandvik}},\ and\
  \citenamefont {{Unsal}}}]{deconfwall}%
  \BibitemOpen
  \bibfield  {author} {\bibinfo {author} {\bibfnamefont {T.}~\bibnamefont
  {{Sulejmanpasic}}}, \bibinfo {author} {\bibfnamefont {H.}~\bibnamefont
  {{Shao}}}, \bibinfo {author} {\bibfnamefont {A.~W.}\ \bibnamefont
  {{Sandvik}}}, \ and\ \bibinfo {author} {\bibfnamefont {M.}~\bibnamefont
  {{Unsal}}},\ }\href@noop {} {\ }\Eprint {http://arxiv.org/abs/1608.09011}
  {arXiv:1608.09011} \BibitemShut {NoStop}%
\bibitem [{\citenamefont {Haldane}(1982)}]{haldane1982}%
  \BibitemOpen
  \bibfield  {author} {\bibinfo {author} {\bibfnamefont {F.~D.~M.}\
  \bibnamefont {Haldane}},\ }\href {\doibase 10.1103/PhysRevB.25.4925}
  {\bibfield  {journal} {\bibinfo  {journal} {Phys. Rev. B}\ }\textbf {\bibinfo
  {volume} {25}},\ \bibinfo {pages} {4925} (\bibinfo {year}
  {1982})}\BibitemShut {NoStop}%
\bibitem [{\citenamefont {Majumdar}(1970)}]{majumdar1970}%
  \BibitemOpen
  \bibfield  {author} {\bibinfo {author} {\bibfnamefont {C.~K.}\ \bibnamefont
  {Majumdar}},\ }\href {http://stacks.iop.org/0022-3719/3/i=4/a=019} {\bibfield
   {journal} {\bibinfo  {journal} {J. Phys. C}\ }\textbf {\bibinfo {volume}
  {3}},\ \bibinfo {pages} {911} (\bibinfo {year} {1970})}\BibitemShut {NoStop}%
\bibitem [{\citenamefont {Singh}\ \emph {et~al.}(1989)\citenamefont {Singh},
  \citenamefont {Fisher},\ and\ \citenamefont {Shankar}}]{singh1989}%
  \BibitemOpen
  \bibfield  {author} {\bibinfo {author} {\bibfnamefont {R.~R.~P.}\
  \bibnamefont {Singh}}, \bibinfo {author} {\bibfnamefont {M.~E.}\ \bibnamefont
  {Fisher}}, \ and\ \bibinfo {author} {\bibfnamefont {R.}~\bibnamefont
  {Shankar}},\ }\href@noop {} {\bibfield  {journal} {\bibinfo  {journal} {Phys.
  Rev. B}\ }\textbf {\bibinfo {volume} {39}},\ \bibinfo {pages} {2562}
  (\bibinfo {year} {1989})}\BibitemShut {NoStop}%
\bibitem [{\citenamefont {Mourigal}\ \emph {et~al.}(2013)\citenamefont
  {Mourigal}, \citenamefont {Enderle}, \citenamefont {Kl{\"o}pperpieper},
  \citenamefont {Caux}, \citenamefont {Stunault},\ and\ \citenamefont
  {R{\o}nnow}}]{mourigal2013}%
  \BibitemOpen
  \bibfield  {author} {\bibinfo {author} {\bibfnamefont {M.}~\bibnamefont
  {Mourigal}}, \bibinfo {author} {\bibfnamefont {M.}~\bibnamefont {Enderle}},
  \bibinfo {author} {\bibfnamefont {A.}~\bibnamefont {Kl{\"o}pperpieper}},
  \bibinfo {author} {\bibfnamefont {J.-S.}\ \bibnamefont {Caux}}, \bibinfo
  {author} {\bibfnamefont {A.}~\bibnamefont {Stunault}}, \ and\ \bibinfo
  {author} {\bibfnamefont {H.~M.}\ \bibnamefont {R{\o}nnow}},\ }\href@noop {}
  {\bibfield  {journal} {\bibinfo  {journal} {Nature Phys.}\ }\textbf {\bibinfo
  {volume} {9}},\ \bibinfo {pages} {435} (\bibinfo {year} {2013})}\BibitemShut
  {NoStop}%
\bibitem [{\citenamefont {Capponni}(2017)}]{sylvain2017}%
  \BibitemOpen
  \bibfield  {author} {\bibinfo {author} {\bibfnamefont {S.}~\bibnamefont
  {Capponni}},\ }\href@noop {} {}\bibinfo {howpublished} {private
  communication} (\bibinfo {year} {2017})\BibitemShut {NoStop}%
\bibitem [{\citenamefont {Heidrich-Meisner}\ \emph {et~al.}(2006)\citenamefont
  {Heidrich-Meisner}, \citenamefont {Honecker},\ and\ \citenamefont
  {Vekua}}]{honecker2006}%
  \BibitemOpen
  \bibfield  {author} {\bibinfo {author} {\bibfnamefont {F.}~\bibnamefont
  {Heidrich-Meisner}}, \bibinfo {author} {\bibfnamefont {A.}~\bibnamefont
  {Honecker}}, \ and\ \bibinfo {author} {\bibfnamefont {T.}~\bibnamefont
  {Vekua}},\ }\href {\doibase 10.1103/PhysRevB.74.020403} {\bibfield  {journal}
  {\bibinfo  {journal} {Phys. Rev. B}\ }\textbf {\bibinfo {volume} {74}},\
  \bibinfo {pages} {020403} (\bibinfo {year} {2006})}\BibitemShut {NoStop}%
\bibitem [{\citenamefont {Schnack}\ \emph {et~al.}(2001)\citenamefont
  {Schnack}, \citenamefont {Schmidt}, \citenamefont {Richter},\ and\
  \citenamefont {Schulenburg}}]{schnack2001}%
  \BibitemOpen
  \bibfield  {author} {\bibinfo {author} {\bibfnamefont {J.}~\bibnamefont
  {Schnack}}, \bibinfo {author} {\bibfnamefont {H.-J.}\ \bibnamefont
  {Schmidt}}, \bibinfo {author} {\bibfnamefont {J.}~\bibnamefont {Richter}}, \
  and\ \bibinfo {author} {\bibfnamefont {J.}~\bibnamefont {Schulenburg}},\
  }\href {\doibase 10.1007/s10051-001-8701-6} {\bibfield  {journal} {\bibinfo
  {journal} {Eur. Phys. J. B}\ }\textbf {\bibinfo {volume} {24}},\ \bibinfo
  {pages} {475} (\bibinfo {year} {2001})}\BibitemShut {NoStop}%
\bibitem [{\citenamefont {Schulenburg}\ \emph {et~al.}(2002)\citenamefont
  {Schulenburg}, \citenamefont {Honecker}, \citenamefont {Schnack},
  \citenamefont {Richter},\ and\ \citenamefont {Schmidt}}]{schulenburg2002}%
  \BibitemOpen
  \bibfield  {author} {\bibinfo {author} {\bibfnamefont {J.}~\bibnamefont
  {Schulenburg}}, \bibinfo {author} {\bibfnamefont {A.}~\bibnamefont
  {Honecker}}, \bibinfo {author} {\bibfnamefont {J.}~\bibnamefont {Schnack}},
  \bibinfo {author} {\bibfnamefont {J.}~\bibnamefont {Richter}}, \ and\
  \bibinfo {author} {\bibfnamefont {H.-J.}\ \bibnamefont {Schmidt}},\ }\href
  {\doibase 10.1103/PhysRevLett.88.167207} {\bibfield  {journal} {\bibinfo
  {journal} {Phys. Rev. Lett.}\ }\textbf {\bibinfo {volume} {88}},\ \bibinfo
  {pages} {167207} (\bibinfo {year} {2002})}\BibitemShut {NoStop}%
\bibitem [{\citenamefont {Richter}\ \emph {et~al.}(2004)\citenamefont
  {Richter}, \citenamefont {Schulenburg}, \citenamefont {Honecker},
  \citenamefont {Schnack},\ and\ \citenamefont {Schmidt}}]{richter2004}%
  \BibitemOpen
  \bibfield  {author} {\bibinfo {author} {\bibfnamefont {J.}~\bibnamefont
  {Richter}}, \bibinfo {author} {\bibfnamefont {J.}~\bibnamefont
  {Schulenburg}}, \bibinfo {author} {\bibfnamefont {A.}~\bibnamefont
  {Honecker}}, \bibinfo {author} {\bibfnamefont {J.}~\bibnamefont {Schnack}}, \
  and\ \bibinfo {author} {\bibfnamefont {H.-J.}\ \bibnamefont {Schmidt}},\
  }\href {http://stacks.iop.org/0953-8984/16/i=11/a=029} {\bibfield  {journal}
  {\bibinfo  {journal} {J. Phys.: Condens. Matter}\ }\textbf {\bibinfo {volume}
  {16}},\ \bibinfo {pages} {S779} (\bibinfo {year} {2004})}\BibitemShut
  {NoStop}%
\bibitem [{\citenamefont {Honecker}\ \emph {et~al.}(2004)\citenamefont
  {Honecker}, \citenamefont {Schulenburg},\ and\ \citenamefont
  {Richter}}]{honecker2004}%
  \BibitemOpen
  \bibfield  {author} {\bibinfo {author} {\bibfnamefont {A.}~\bibnamefont
  {Honecker}}, \bibinfo {author} {\bibfnamefont {J.}~\bibnamefont
  {Schulenburg}}, \ and\ \bibinfo {author} {\bibfnamefont {J.}~\bibnamefont
  {Richter}},\ }\href {http://stacks.iop.org/0953-8984/16/i=11/a=025}
  {\bibfield  {journal} {\bibinfo  {journal} {J. Phys.: Condens. Matter}\
  }\textbf {\bibinfo {volume} {16}},\ \bibinfo {pages} {S749} (\bibinfo {year}
  {2004})}\BibitemShut {NoStop}%
\bibitem [{\citenamefont {Balents}\ and\ \citenamefont
  {Starykh}(2016)}]{balents2015}%
  \BibitemOpen
  \bibfield  {author} {\bibinfo {author} {\bibfnamefont {L.}~\bibnamefont
  {Balents}}\ and\ \bibinfo {author} {\bibfnamefont {O.~A.}\ \bibnamefont
  {Starykh}},\ }\href {\doibase 10.1103/PhysRevLett.116.177201} {\bibfield
  {journal} {\bibinfo  {journal} {Phys. Rev. Lett.}\ }\textbf {\bibinfo
  {volume} {116}},\ \bibinfo {pages} {177201} (\bibinfo {year}
  {2016})}\BibitemShut {NoStop}%
\bibitem [{\citenamefont {Oshikawa}\ \emph {et~al.}(1997)\citenamefont
  {Oshikawa}, \citenamefont {Yamanaka},\ and\ \citenamefont
  {Affleck}}]{oshikawa1997}%
  \BibitemOpen
  \bibfield  {author} {\bibinfo {author} {\bibfnamefont {M.}~\bibnamefont
  {Oshikawa}}, \bibinfo {author} {\bibfnamefont {M.}~\bibnamefont {Yamanaka}},
  \ and\ \bibinfo {author} {\bibfnamefont {I.}~\bibnamefont {Affleck}},\ }\href
  {\doibase 10.1103/PhysRevLett.78.1984} {\bibfield  {journal} {\bibinfo
  {journal} {Phys. Rev. Lett.}\ }\textbf {\bibinfo {volume} {78}},\ \bibinfo
  {pages} {1984} (\bibinfo {year} {1997})}\BibitemShut {NoStop}%
\bibitem [{\citenamefont {Haldane}(1981)}]{haldane1981}%
  \BibitemOpen
  \bibfield  {author} {\bibinfo {author} {\bibfnamefont {F.~D.~M.}\
  \bibnamefont {Haldane}},\ }\href {\doibase 10.1103/PhysRevLett.47.1840}
  {\bibfield  {journal} {\bibinfo  {journal} {Phys. Rev. Lett.}\ }\textbf
  {\bibinfo {volume} {47}},\ \bibinfo {pages} {1840} (\bibinfo {year}
  {1981})}\BibitemShut {NoStop}%
\bibitem [{\citenamefont {Tonegawa}\ and\ \citenamefont
  {Harada}(1987)}]{tonegawa1987}%
  \BibitemOpen
  \bibfield  {author} {\bibinfo {author} {\bibfnamefont {T.}~\bibnamefont
  {Tonegawa}}\ and\ \bibinfo {author} {\bibfnamefont {I.}~\bibnamefont
  {Harada}},\ }\href {\doibase 10.1143/JPSJ.56.2153} {\bibfield  {journal}
  {\bibinfo  {journal} {J. Phys. Soc. Jpn.}\ }\textbf {\bibinfo {volume}
  {56}},\ \bibinfo {pages} {2153} (\bibinfo {year} {1987})}\BibitemShut
  {NoStop}%
\bibitem [{\citenamefont {Farnell}\ and\ \citenamefont
  {Parkinson}(1994)}]{farnell1993}%
  \BibitemOpen
  \bibfield  {author} {\bibinfo {author} {\bibfnamefont {D.~J.~J.}\
  \bibnamefont {Farnell}}\ and\ \bibinfo {author} {\bibfnamefont {J.~B.}\
  \bibnamefont {Parkinson}},\ }\href
  {http://stacks.iop.org/0953-8984/6/i=28/a=024} {\bibfield  {journal}
  {\bibinfo  {journal} {J. Phys.: Condens. Matter}\ }\textbf {\bibinfo {volume}
  {6}},\ \bibinfo {pages} {5521} (\bibinfo {year} {1994})}\BibitemShut
  {NoStop}%
\bibitem [{\citenamefont {Gerhardt}\ \emph {et~al.}(1997)\citenamefont
  {Gerhardt}, \citenamefont {Fledderjohann}, \citenamefont {Aysal},
  \citenamefont {MŸtter}, \citenamefont {Audet},\ and\ \citenamefont
  {Kršger}}]{gerhardt1997}%
  \BibitemOpen
  \bibfield  {author} {\bibinfo {author} {\bibfnamefont {C.}~\bibnamefont
  {Gerhardt}}, \bibinfo {author} {\bibfnamefont {A.}~\bibnamefont
  {Fledderjohann}}, \bibinfo {author} {\bibfnamefont {E.}~\bibnamefont
  {Aysal}}, \bibinfo {author} {\bibfnamefont {K.-H.}\ \bibnamefont {MŸtter}},
  \bibinfo {author} {\bibfnamefont {J.~F.}\ \bibnamefont {Audet}}, \ and\
  \bibinfo {author} {\bibfnamefont {H.}~\bibnamefont {Kršger}},\ }\href
  {http://stacks.iop.org/0953-8984/9/i=16/a=015} {\bibfield  {journal}
  {\bibinfo  {journal} {J. Phys.: Condens. Matter}\ }\textbf {\bibinfo {volume}
  {9}},\ \bibinfo {pages} {3435} (\bibinfo {year} {1997})}\BibitemShut
  {NoStop}%
\bibitem [{\citenamefont {Lu}\ \emph {et~al.}(2006)\citenamefont {Lu},
  \citenamefont {Wang}, \citenamefont {Qin},\ and\ \citenamefont
  {Xiang}}]{lu2006}%
  \BibitemOpen
  \bibfield  {author} {\bibinfo {author} {\bibfnamefont {H.~T.}\ \bibnamefont
  {Lu}}, \bibinfo {author} {\bibfnamefont {Y.~J.}\ \bibnamefont {Wang}},
  \bibinfo {author} {\bibfnamefont {S.}~\bibnamefont {Qin}}, \ and\ \bibinfo
  {author} {\bibfnamefont {T.}~\bibnamefont {Xiang}},\ }\href {\doibase
  10.1103/PhysRevB.74.134425} {\bibfield  {journal} {\bibinfo  {journal} {Phys.
  Rev. B}\ }\textbf {\bibinfo {volume} {74}},\ \bibinfo {pages} {134425}
  (\bibinfo {year} {2006})}\BibitemShut {NoStop}%
\bibitem [{\citenamefont {{He}}\ \emph {et~al.}()\citenamefont {{He}},
  \citenamefont {{Jiang}}, \citenamefont {{Yu}}, \citenamefont {{Lin}},\ and\
  \citenamefont {{Guan}}}]{he2017}%
  \BibitemOpen
  \bibfield  {author} {\bibinfo {author} {\bibfnamefont {F.}~\bibnamefont
  {{He}}}, \bibinfo {author} {\bibfnamefont {Y.-Z.}\ \bibnamefont {{Jiang}}},
  \bibinfo {author} {\bibfnamefont {Y.-C.}\ \bibnamefont {{Yu}}}, \bibinfo
  {author} {\bibfnamefont {H.-Q.}\ \bibnamefont {{Lin}}}, \ and\ \bibinfo
  {author} {\bibfnamefont {X.-W.}\ \bibnamefont {{Guan}}},\ }\href@noop {} {\
  }\Eprint {http://arxiv.org/abs/1702.05903} {arXiv:1702.05903} \BibitemShut
  {NoStop}%
\bibitem [{\citenamefont {Kono}\ \emph {et~al.}(2015)\citenamefont {Kono},
  \citenamefont {Sakakibara}, \citenamefont {Aoyama}, \citenamefont {Hotta},
  \citenamefont {Turnbull}, \citenamefont {Landee},\ and\ \citenamefont
  {Takano}}]{kono2015}%
  \BibitemOpen
  \bibfield  {author} {\bibinfo {author} {\bibfnamefont {Y.}~\bibnamefont
  {Kono}}, \bibinfo {author} {\bibfnamefont {T.}~\bibnamefont {Sakakibara}},
  \bibinfo {author} {\bibfnamefont {C.~P.}\ \bibnamefont {Aoyama}}, \bibinfo
  {author} {\bibfnamefont {C.}~\bibnamefont {Hotta}}, \bibinfo {author}
  {\bibfnamefont {M.~M.}\ \bibnamefont {Turnbull}}, \bibinfo {author}
  {\bibfnamefont {C.~P.}\ \bibnamefont {Landee}}, \ and\ \bibinfo {author}
  {\bibfnamefont {Y.}~\bibnamefont {Takano}},\ }\href {\doibase
  10.1103/PhysRevLett.114.037202} {\bibfield  {journal} {\bibinfo  {journal}
  {Phys. Rev. Lett.}\ }\textbf {\bibinfo {volume} {114}},\ \bibinfo {pages}
  {037202} (\bibinfo {year} {2015})}\BibitemShut {NoStop}%
\bibitem [{\citenamefont {Jeong}\ and\ \citenamefont
  {R\o{}nnow}(2015)}]{jeong2015}%
  \BibitemOpen
  \bibfield  {author} {\bibinfo {author} {\bibfnamefont {M.}~\bibnamefont
  {Jeong}}\ and\ \bibinfo {author} {\bibfnamefont {H.~M.}\ \bibnamefont
  {R\o{}nnow}},\ }\href {\doibase 10.1103/PhysRevB.92.180409} {\bibfield
  {journal} {\bibinfo  {journal} {Phys. Rev. B}\ }\textbf {\bibinfo {volume}
  {92}},\ \bibinfo {pages} {180409} (\bibinfo {year} {2015})}\BibitemShut
  {NoStop}%
\bibitem [{\citenamefont {Watson}\ \emph {et~al.}(2001)\citenamefont {Watson},
  \citenamefont {Kotov}, \citenamefont {Meisel}, \citenamefont {Hall},
  \citenamefont {Granroth}, \citenamefont {Montfrooij}, \citenamefont {Nagler},
  \citenamefont {Jensen}, \citenamefont {Backov}, \citenamefont {Petruska},
  \citenamefont {Fanucci},\ and\ \citenamefont {Talham}}]{watson2001}%
  \BibitemOpen
  \bibfield  {author} {\bibinfo {author} {\bibfnamefont {B.~C.}\ \bibnamefont
  {Watson}}, \bibinfo {author} {\bibfnamefont {V.~N.}\ \bibnamefont {Kotov}},
  \bibinfo {author} {\bibfnamefont {M.~W.}\ \bibnamefont {Meisel}}, \bibinfo
  {author} {\bibfnamefont {D.~W.}\ \bibnamefont {Hall}}, \bibinfo {author}
  {\bibfnamefont {G.~E.}\ \bibnamefont {Granroth}}, \bibinfo {author}
  {\bibfnamefont {W.~T.}\ \bibnamefont {Montfrooij}}, \bibinfo {author}
  {\bibfnamefont {S.~E.}\ \bibnamefont {Nagler}}, \bibinfo {author}
  {\bibfnamefont {D.~A.}\ \bibnamefont {Jensen}}, \bibinfo {author}
  {\bibfnamefont {R.}~\bibnamefont {Backov}}, \bibinfo {author} {\bibfnamefont
  {M.~A.}\ \bibnamefont {Petruska}}, \bibinfo {author} {\bibfnamefont {G.~E.}\
  \bibnamefont {Fanucci}}, \ and\ \bibinfo {author} {\bibfnamefont {D.~R.}\
  \bibnamefont {Talham}},\ }\href {\doibase 10.1103/PhysRevLett.86.5168}
  {\bibfield  {journal} {\bibinfo  {journal} {Phys. Rev. Lett.}\ }\textbf
  {\bibinfo {volume} {86}},\ \bibinfo {pages} {5168} (\bibinfo {year}
  {2001})}\BibitemShut {NoStop}%
\bibitem [{\citenamefont {{Mao}}\ \emph {et~al.}()\citenamefont {{Mao}},
  \citenamefont {{Cheng}}, \citenamefont {{Chen}},\ and\ \citenamefont
  {{Luo}}}]{mao2017}%
  \BibitemOpen
  \bibfield  {author} {\bibinfo {author} {\bibfnamefont {B.-B.}\ \bibnamefont
  {{Mao}}}, \bibinfo {author} {\bibfnamefont {C.}~\bibnamefont {{Cheng}}},
  \bibinfo {author} {\bibfnamefont {F.-Z.}\ \bibnamefont {{Chen}}}, \ and\
  \bibinfo {author} {\bibfnamefont {H.-G.}\ \bibnamefont {{Luo}}},\ }\href@noop
  {} {\ }\Eprint {http://arxiv.org/abs/1701.00013} {arXiv:1701.00013}
  \BibitemShut {NoStop}%
\bibitem [{\citenamefont {Suwa}(2014)}]{suwa2014}%
  \BibitemOpen
  \bibfield  {author} {\bibinfo {author} {\bibfnamefont {H.}~\bibnamefont
  {Suwa}},\ }\href {\doibase 10.1007/978-4-431-54517-0} {\emph {\bibinfo
  {title} {Geometrically Constructed Markov Chain Monte Carlo Study of Quantum
  Spin-phonon Complex Systems}}}\ (\bibinfo  {publisher} {Springer Japan},\
  \bibinfo {year} {2014})\ Chap.~\bibinfo {chapter} {6}\BibitemShut {NoStop}%
\bibitem [{\citenamefont {Sandvik}\ and\ \citenamefont
  {Campbell}(1999)}]{sandvik1999}%
  \BibitemOpen
  \bibfield  {author} {\bibinfo {author} {\bibfnamefont {A.~W.}\ \bibnamefont
  {Sandvik}}\ and\ \bibinfo {author} {\bibfnamefont {D.~K.}\ \bibnamefont
  {Campbell}},\ }\href {\doibase 10.1103/PhysRevLett.83.195} {\bibfield
  {journal} {\bibinfo  {journal} {Phys. Rev. Lett.}\ }\textbf {\bibinfo
  {volume} {83}},\ \bibinfo {pages} {195} (\bibinfo {year} {1999})}\BibitemShut
  {NoStop}%
\end{thebibliography}%

\end{document}